\UseRawInputEncoding
\documentclass[journal,12pt,onecolumn,draftclsnofoot,]{IEEEtran}
\usepackage[bookmarks=false]{hyperref}
\IEEEoverridecommandlockouts
\usepackage[noadjust]{cite}
\usepackage[mathscr]{euscript}
\usepackage[noend]{algpseudocode}
\usepackage{array}
\usepackage{mdwmath}
\usepackage{mathtools}
\usepackage{float} 
\usepackage{eqparbox}
\usepackage[ruled,vlined]{algorithm2e}
\usepackage{mdframed}
\usepackage{steinmetz}
\usepackage{stfloats}
\usepackage{graphicx}
\usepackage{url}
\usepackage{subcaption}
\graphicspath{{../pdf/}{../jpeg/}}
\DeclareGraphicsExtensions{.pdf,.jpeg,.png}
\usepackage{epstopdf}
\usepackage{amsfonts}
\newtheorem{lemma}{Lemma}
\newtheorem{theorem}{Theorem}

\newtheorem{corollary}{Corollary}

\usepackage{latexsym}
\usepackage{amsfonts,amssymb,amsmath}
\usepackage{hyperref}
\begin{document}

\title{Performance Analysis under IRS-User Association for Distributed IRSs Assisted MISO Systems} 
\author{Hibatallah Alwazani, \IEEEmembership{Student Member, IEEE}, Qurrat-Ul-Ain Nadeem, \IEEEmembership{Member, IEEE},  Anas Chaaban, \IEEEmembership{Senior Member, IEEE}\thanks{H. Alwazani, Q.-U.-A. Nadeem, and A. Chaaban are with the School of Engineering, University of British Columbia, Kelowna, BC V1V1V7, Canada. (email: \{hibat97, qurrat.nadeem,anas.chaaban\}@ubc.ca).}
\thanks{
This publication is based upon work supported by the King Abdullah University of Science and Technology (KAUST) under Award No. OSR-2018-CRG7-3734.
%Parts of the results in this paper were presented in %xxx \cite{xxx}.
}}\maketitle
\begin{abstract}
Distributed intelligent reflecting surfaces (IRSs) deployed in multi-user wireless communication systems promise improved system performance. However, the signal-to-interference-plus-noise ratio (SINR) analysis and IRSs optimization in such a system become challenging, due to the large number of involved parameters. The system optimization can be simplified if users are associated with IRSs, which in turn focus on serving the associated users. We provide a practical theoretical framework for the average SINR analysis of a distributed IRSs-assisted multi-user MISO system, where IRSs are optimized to serve their associated users. In particular, we derive the average SINR expression under maximum ratio transmission (MRT) precoding at the BS and optimized reflect beamforming configurations at the IRSs. A successive refinement (SR) method is then outlined to optimize the IRS-user association parameters for the formulated max-min SINR problem which motivates user-fairness. Simulations validate the average SINR analysis while confirming the superiority of a distributed IRSs system over a centralized IRS system as well as the gains with optimized IRS-user association as compared to random association.
\end{abstract}
\section{Introduction}
\label{intro}
\IEEEPARstart{W}{ith} the imminence of massive connectivity promising a plethora of devices to be able to communicate effectively comes a new set of problems. How to create  robust, high-speed links sustained over a wide range of geographical locations for many devices? Smart radio environments, where the environment is jointly optimized along with the transmitter and receiver, is a promising concept to solve these problems  \cite{smartRadio}. To enable the control of radio environments, communication engineers are exploring the idea of deploying  software-controlled  surfaces, referred to as intelligent reflecting surfaces (IRSs), on structures in the environment such as buildings. Each IRS contains a large number of low-cost, passive reflecting elements, where each element can introduce a phase shift onto the impinging electromagnetic (EM) waves  to achieve a desired objective. This objective can be in the form of increasing the received power of the desired signal at a receiver, or decreasing the power of interference, or increasing the ratio between the two. An IRS can help to achieve these objectives via  beam-focusing, which is tuning the magnitude of the radiation pattern in a certain direction, and/or beam-steering which is modifying the direction of the beam. 

Designing IRS phase shifts to shape the impinging EM waves, referred to as passive (or reflect) beamforming, has been studied  extensively in recent literature \cite{maxmin,practicalPhase,distbeamforming,energyBeamforming}.  The works in \cite{maxmin}, \cite{ practicalPhase}, and \cite{ energyBeamforming} focus on jointly optimizing transmit beamforming at the BS and passive beamforming at the IRS to meet a certain goal such as maximizing energy efficiency \cite{energyBeamforming}, maximizing minimum user SINR subject to a transmit power constraint \cite{maxmin}, or minimizing transmit power subject to quality of service constraints \cite{practicalPhase}. IRSs have also been studied to enhance physical layer security in \cite{covertcomm}.  %Furthermore, the authors in \cite{nomavsoma} investigate the theoretical performance comparison between  orthogonal multiple access (OMA) and non-orthogonal multiple access (NOMA) in an IRS-assisted downlink communication system.
For a general setting with frequency selective channels, an IRS-aided orthogonal frequency division multiplexing (OFDM) based wireless system is tackled in \cite{ofdm1} and \cite{ofdm2}. Moreover, IRSs have  found applications in simultaneous wireless information and power transfer (SWIPT) \cite{swipt}, \cite{swipt2}. All these works consider a single IRS in their system model.%the authors  in  provide novel theoretical foundations study under an IRS-aided SWIPT system, likewise \cite{swipt2} unveils the impact of IRS on transmission design for SWIPT.

% Distributed IRSs is a promising solution to provide seamless coverage to a large area and solve the rank-deficiency problem that can be caused by a single LoS IRS. In a system with distributed IRSs, the overall BS-IRS link will be the sum of multiple low rank channels which guarantees high rank channels even when the IRSs are deployed in the LoS of the BS \cite{beamforminddistributed}. 
The current  literature on distributed IRSs focus on the design of reflect beamforming to increase coverage and performance \cite{kishkds1, beamforminddistributed}. The authors in \cite{kishkds1} study the effects of large-scale deployment of IRSs in cellular communication by equipping blocking structures with IRSs and eliminating blind-spots. % Adding a single IRS with a sufficient number of elements can provide extended coverage, which is usually limited to the vicinity of the IRS. 
Generally, deploying a single IRS in the line of sight (LoS) of the BS can reduce the degrees of freedom of the overall channel to one (or a low number as compared to the number of BS antennas and IRS elements) \cite{nadeem}. The resulting rank deficient system can not serve multiple users simultaneously. In \cite{beamforminddistributed}, distributed IRSs are explored as a promising solution to the rank-deficiency problem because the overall BS-IRS channel in the distributed IRSs case would be the sum of multiple rank one channels which guarantees higher
rank channels. The goal in \cite{beamforminddistributed} is to maximize sum-rate via  jointly optimizing the transmit power and the phase shift matrices at all the distributed IRSs. The authors in \cite{IRSuserassociation}  handle the IRS-user association problem by assigning IRSs to users to optimally balance the passive beamforming gains among different BS-user links. Moreover, they derive the signal-to-interference-plus-noise ratio (SINR) expression at each user in closed form, and create an SINR balancing optimization problem to find the IRS-user association parameters. However, their system model assumes single-antenna BSs to simplify the analysis.

In this work, we formulate and solve a max-min average SINR optimization problem for a distributed IRSs assisted multi-user MISO system to find the optimal IRS-user association. We focus on the average (ergodic) analysis of the SINR at each user, under the scenario where each IRS is associated with one user in the system. For a particular IRS-user association pair, we choose the design for IRS phase shifts that would maximize the received signal strength at that user. Under that design, we utilize statistical tools to obtain a closed form expression for the average SINR at each user. Finally, we outline a low-complexity SR algorithm to find the IRS-user association parameters that maximize the minimum SINR subject to constraints on the values of binary association parameters. The optimization of IRS-user association in a distributed IRSs setting has been rarely dealt with in literature, and as more research expands in IRS-aided systems, a natural outcome is to have IRSs deployed in different geographical locations associated with different users or groups of users, in order to sustain and enable the concept of smart radio environments. In particular, the existing work on IRS-user association \cite{IRSuserassociation}   simplifies the system model by assuming  single-antenna BSs, while in our work we consider a multiple-antenna BS which is  practical.

The results illustrate the significance of associating the IRSs with users in an optimized fashion resulting in an increase in the SINR of the bottleneck user, when compared to settings where  the IRSs are randomly assigned to the users or where the IRSs are assigned based on the minimum distance to the users.  The results also showcase the close performance of SR algorithm  to the optimal but computationally expensive exhaustive search in finding the association parameters. The performance improvement by distributing the IRSs in different geographical locations instead of having a centralized IRS unit is also illustrated.

This paper is organized as follows. Sec. \ref{chsysmod}  introduces the distributed IRSs assisted multi-user MISO system model. Sec. \ref{secavg} performs the ergodic SINR analysis for all users and Sec. \ref{secoptimizedIRS} solves the IRS-user association problem with max-min average SINR as an objective. Numerical evaluations and discussions on performance are provided in Sec. \ref{secresults}. Finally, Sec. \ref{chconc} concludes the paper and highlights future directions.

 \textit{Notation:} This notation is used throughout the work. Bold  lower-case and upper-case characters denote vectors and matrices, respectively. The subscripts $(\cdot)^{T}$ and $(\cdot)^{H}$ represent the transpose and Hermitian, respectively. The operator $tr(\cdot)$ is the trace of matrix, while the $\mathbb{E}[\cdot]$ and $\text{Var}[\cdot]$ are the expectation and variance of a random variable, respectively.  The Kronecker and Hadamard products are denoted as $\otimes$ and $\odot$, respectively. For a complex number $x$, define its conjugate as $x^*$,  $|x|$ as its magnitude,  $\angle x$ as its phase (for vectors, the magnitude and phase are taken element-wise). The Euclidean norm of a vector is defined as $\|\cdot\|$. For a diagonal matrix $\mathbf{D}$, $\text{diag}(\mathbf{D})$ refers to its vectored form, while for a a vector $\mathbf{v}$, $\text{diag}(\mathbf{v})$ creates a diagonal matrix. An $M \times M$ identity matrix is denoted by $\mathbf{I}_{M}$, moreover  $\mathbf{1}_{M}$ defines an $M \times M$ all-ones matrix.
 
\begin{figure}
    \centering
    \includegraphics[scale=0.3]{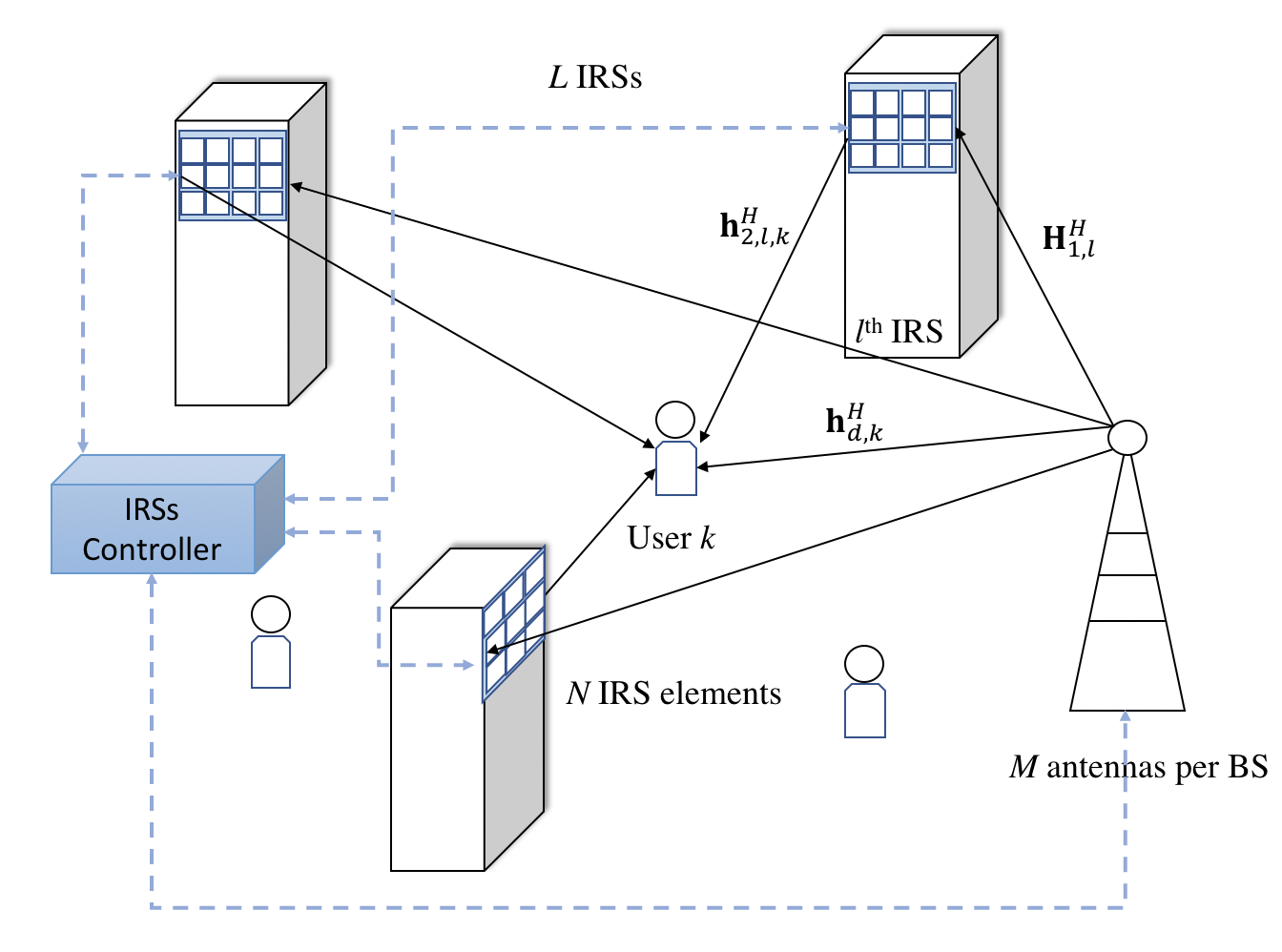}
    \caption{Distributed IRSs assisted MISO system model.}
    \label{Sysmodel}
\end{figure}

\section{System Model}
\label{chsysmod}
\label{chprelim}
 Consider an $M$-antenna BS serving $K$ single-antenna users in the downlink, while being assisted by $L$ IRSs each equipped with $N$ elements (see Fig. \ref{Sysmodel}). The IRSs are deployed in the environment in a distributed manner with fixed positions, and their operation is controlled by a centralized IRS controller that communicates with the BS over a backhaul link \cite{maxmin}.  

The channels are modeled as Rayleigh fading for the direct BS-user and IRS-user links, and as LoS for each BS-IRS link. We assume that the BS has perfect channel state information (CSI) to design the precoder and IRS reflect beamforming vectors. While perfect CSI acquisition is challenging in IRS-assisted systems, we make this assumption to enable a tractable theoretical analysis of the distributed IRSs assisted system, which is already  complex given the large number of links involved. 

%\begin{assumption}
%\label{assum2csi}
%The BS has perfect channel state information (CSI) to design the precoder and IRS reflect beamforming vectors.
%\end{assumption}
%%It is well-known that acquiring perfect CSI is at worst unattainable at best challenging. Yet,  
%{Assumption} \ref{assum2csi} is made to ensure a tractable analysis henceforth since the distributed IRSs MISO system is complex where many links are involved. 

\subsection{Downlink Transmission}

The overall channel between the BS and user $k$ is given by
\begin{align}
\label{eqhksys}
&\mathbf{h}_{k}=\mathbf{h}_{d,k}+\sum_{l=1}^L \mathbf{H}_{1,l}\boldsymbol{\Theta}_l\mathbf{h}_{2,l,k},
\end{align}
where  $\mathbf{h}_{d,k}\in \mathbb{C}^{M\times 1}$ is the direct channel between BS and user $k$, $\mathbf{H}_{1,l} \in \mathbb{C}^{M\times N}$ is the channel between BS and IRS $l$, $\mathbf{h}_{2,l,k}\in \mathbb{C}^{N\times 1}$ is the channel between IRS $l$ and user $k$, and $\boldsymbol{\Theta}_l=\text{diag}(\alpha_{l,1} e^{j \theta_{l,1}}, \dots,$ $\alpha_{l,N}e^{j\theta_{l,N}})\in \mathbb{C}^{N\times N}$ is the reflection matrix for IRS $l$, where $\theta_{l,n}\in [0,2\pi]$ is the phase-shift applied by element $n$ of IRS $l$ and $\alpha_{l,n}\in [0,1]$ is the amplitude reflection  coefficient. We assume the reflection coefficients $\alpha_{l,n}$'s to equal one, as done in most current works on IRSs motivated by the significant advancements made in the design of lossless metasurfaces \cite{losslessmetasurfaces}. The analysis can be straightforwardly extended to arbitrary values of $\alpha_{l,n}$s.

%Note that the distributed IRSs assisted MISO system is  a generalization of the single IRS assisted MISO system \cite{nadeem}, therefore, we build on \cite{nadeem} and extend the system model discussed to distributed surfaces while retaining the similar transmission conditions explained next.

%In this distributed IRSs-assisted MISO system model, we consider a time-division duplexing (TDD) transmission mode  due to the prohibitively large system dimensions ergo large numbers of channels to estimate. Here, TDD exploits channel reciprocity and thus eliminates the need to estimate channels both in the uplink and downlink (for example $\mathbf{h}_k^H$  in the downlink is the Hermetian of $\mathbf{h}_k$ defined in \eqref{eqhksys} in the uplink).

The received baseband signal in the downlink  at user $k$ is defined as
\begin{align}
\label{eqtransmit}
    {y}_k=\mathbf{h}_{k}^H\mathbf{x}+{n}_{k},
\end{align}
where $n_k \sim \mathcal{CN}(0, \sigma^2)$ is the noise at the receiver with noise variance $\sigma^2$, and $\mathbf{x} \in \mathbb{C}^{M \times 1}$ is the transmit signal containing information intended for all users. The transmit signal is formulated as
\begin{align}
\label{eqsignal}
&\mathbf{x}=\sum_{k=1}^{K} \sqrt{p_{k}} \mathbf{f}_{k} s_{k},
\end{align}
where $p_k$ and $s_k\in \mathcal{CN}(0,1)$ are the allocated power and data symbol of user $k$ respectively while $\mathbf{f}_k$ is the precoding vector. We consider  MRT precoding at the BS, which is a popular scheme for large scale MIMO systems due to its low computational complexity, robustness, and high asymptotic performance \cite{massive1, massive2}. The precoding vector $ \mathbf{f}_k \in \mathbb{C}^{M \times 1}$ is given by \cite{mrtprecodingexpectation}
\begin{align}
\label{eqfk}
    \mathbf{f}_k=\frac{\mathbf{h}_k}{ \sqrt{\mathbb{E}[\|\mathbf{h}_k\|^2]}}
\end{align}
where $\mathbf{h}_k$ is stated in \eqref{eqhksys}. Recall that we assume the channel $\mathbf{h}_k$ is known at the base station under some channel estimation scheme for an IRS assisted system \cite{alwazani2020channel}. 

%Note that MRT is optimal for single-user systems. Nevertheless, it is shown in \cite{howmanyantennas} that for large system dimensions such as when when $M$ becomes large MRT  reaches the performance of optimal linear precoding due to channel hardening and favorable propagation. Another reason MRT is chosen as the BS precoding technique is because its simpler than its counterparts (for example. ZF precoding). Since this work focuses on IRSs (and not on BS beamforming techniques), we resort to MRT to find tractable expressions.

%The reason we normalize the precoding vector using the square root of average squared norm $\sqrt{\mathbb{E}[\|\mathbf{h}_k\|^2]}$ of the channel instead of the more conventional norm $\|\mathbf{h}_k\|$ is to provide analytic tractability ,  and enable the results given in Sec. \ref{secavg}.

%Thus, 

 Given $s_k$'s are independently and identically distributed (i.i.d.) $\mathcal{CN}(0,1)$ variables, $\mathbf{x}$ has to satisfy the average power constraint $ \mathbb{E}[||\mathbf{x}||^{2}]={tr}(\mathbf{P} \mathbf{F}^{H} \mathbf{F}) \leq  P_{max},$
where $P_{max}>0$ is the power constraint at the BS,  $\mathbf{P}=\text{diag}(p_1, \dots, p_K)\in \mathbb{C}^{K\times K}$ is the power allocation matrix and $\mathbf{F}=[\mathbf{f}_1, \dots, \mathbf{f}_K]\in \mathbb{C}^{M\times K}$ is the precoding matrix.

 In the next subsection, we describe the channel models for all links.

\subsection{Channel Models}

We assume block fading channel model for $\mathbf{h}_{2,l,k}$ and $\mathbf{h}_{d,k}$  given by independent Rayleigh fading  represented as
\begin{align}
\label{model_h2lk}
&\mathbf{h}_{2,l,k}= \sqrt{\beta_{2,l,k}} \mathbf{z}_{l,k}, \\
\label{model_hdk}
&\mathbf{h}_{d,k}=\sqrt{\beta_{d,k}}  \mathbf{z}_{d,k},
\end{align}
where  $\beta_{2,l,k}$ is the path loss factor for the IRS-user $k$ channel and $\beta_{d,k}$  is the path loss factor for the direct channel, and $\mathbf{z}_{l,k} \sim \mathcal{CN}(0,\mathbf{I}_{N}) $ and $\mathbf{z}_{d,k} \sim \mathcal{CN}(0,\mathbf{I}_{M})$ describe the fast fading vectors of the IRS-user channel and the BS-user channel, respectively.

 %Notice that the correlation matrices of the direct and IRS-user channel are set to identity which shifts the model to uncorrelated Rayleigh fading. In practice, channels are spatially correlated, but we force the uncorrelated fading assumption at the BS and IRS to obtain tractable closed-form expression for the average SINR under the umbrella of perfect CSI.  

%\footnote{The analysis is extendible (with some very tedious algebraic manipulations) to correlated Rayleigh fading  and imperfect CSI.}

%In addition, the notion of block fading in this channel modelling indicates that $\mathbf{h}_{2,l,k}$ and $\mathbf{h}_{d,k}$ remain constant during a duration of one block (code length), and change independently between blocks. 

 %Recall there are two channels which make up each IRS cascaded channel, one of which is the BS-IRS $l$ channel $\mathbf{H}_{1,l}$ which is defined here. 

We assume the BS-IRS channels to be LoS dominated as assumed in many  other works on IRS-assisted systems, for example: \cite{los_channel,nadeem2020intelligent,los, maxmin, LIS_los}. This assumption is quite practical and is supported in literature with the following remarks:
\begin{itemize}
    \item First, as the BS tower is generally elevated high and the IRS is also envisioned to be integrated onto the walls of (high-rise) buildings, so both will have few obstacles around. Given the positions of BS and IRSs are fixed, a stable LoS channel between the BS and each IRS will exist and can be constructed at  the BS using directional (LoS angle of departure (AoD) and angle of arrival (AoA)) information. 
    \item Second, the path loss in NLoS paths is much larger than that in the LoS path in the next generation systems due to the transition to higher frequencies. In fact, it is noted that in mmWave systems  the typical value of Rician factor (ratio of energy in LoS component to that in NLoS components) is $20\rm{dB}$ and can be as large as $40$\rm{dB} in some cases \cite{los_channel}, which is sufficiently large to neglect any NLoS channel components. 
\end{itemize}

Given their LoS nature, the BS can construct the BS-IRS channels and thus obtain prior knowledge of $\mathbf{H}_{1,l},\forall l$.%, making this channel deterministic.
\begin{figure}
    \centering
    \includegraphics[scale=0.45]{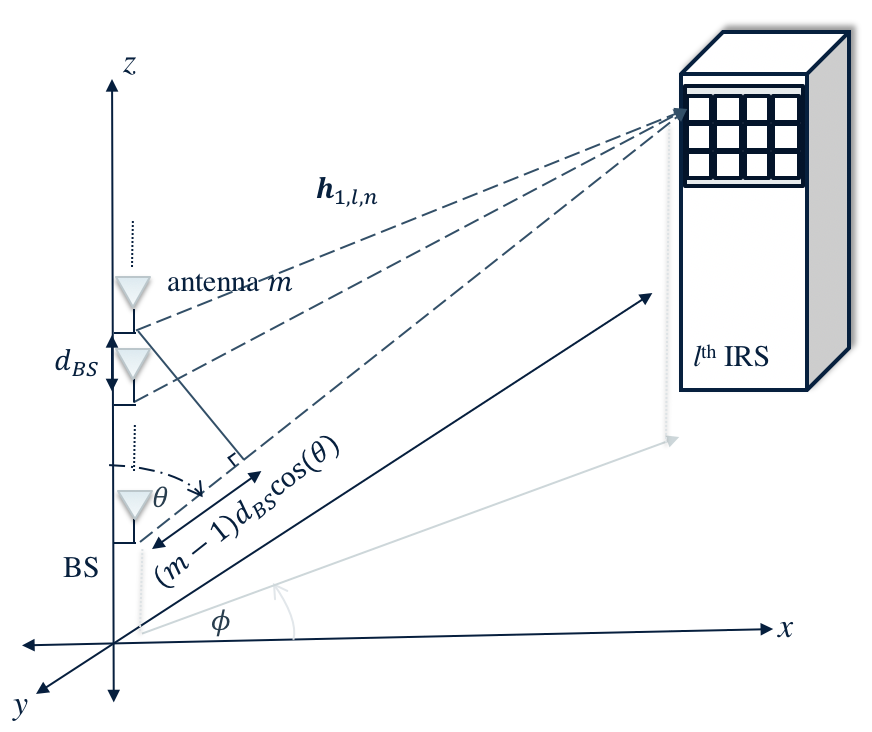}
    \caption[LoS BS-IRS $l$ channel $\mathbf{H}_{1,l}$ visualization]{{ Model of $\mathbf{h}_{1,l,n} \in \mathbb{C}^{M \times 1}$ which is the $n^{th}$ column vector of of LoS BS-IRS $l$ channel matrix $\mathbf{H}_{1,l}$. The signals from transmit antennas arrive almost in parallel at the $n^{th}$ IRS element. }} 
    \label{fig:h1l}
\end{figure}

    %{images/ceimages/updatedbschannel.png}

Assuming a uniform linear array (ULA) at the BS and uniform planar array (UPA) at the IRSs depicted in Fig. \ref{fig:h1l}, the LoS BS-IRS $l$ channel can be written as
\begin{align}
\label{eqmodel_H1}
&\mathbf{H_{1,l}}=  \sqrt{\beta_{1,l}}\mathbf{a}_l\mathbf{b}_l^H.
\end{align}
Here, $\beta_{1,l}$ is the path loss factor for BS-IRS $l$ channel,  $\mathbf{a}_l$ is the array response vector for the BS, while $\mathbf{b}_l$ is the array response vector for IRS $l$. 
The  vector $\mathbf{a}_l \in \mathbb{C}^{M \times 1}$ is written as \cite{tsewirelesscomm}
\begin{align}
    \mathbf{a}_l=[1, e^{-j kd_{BS}\cos(\theta_l)}, \dots, e^{-j kd_{BS}(M-1)\cos(\theta_l)}]^T,
\end{align}
where $d_{BS}$ is the inter-antenna spacing, $k=2 \pi/\lambda_c$ is the wave number,  $\lambda_c$ is carrier wavelength, $\theta_l$ is the elevation AoD from the BS to IRS $l$ (cf. Fig.\ref{fig:h1l}).  %  $\theta$ is interpreted as the angle made by the LoS path from the BS to IRS $l$ with the boresite of the ULA at the BS. 

The IRS is envisioned to be a planar array of $N=N_xN_z$ elements, where $N_x$ and $N_z$ denote the number of horizontally and vertically placed elements, respectively. The array response vector $   \mathbf{b}_l \in \mathbb{C}^{N \times 1}$ for a UPA at IRS $l$ is expressed as 
to be
   $ \mathbf{b}_l=\mathbf{b}_{l,x}^T
    \otimes \mathbf{b}_{l,z}^T$ and \cite{upachannel1}
    \begin{align}
   &\mathbf{b}_{l,x}= [1, e^{-j kd_{IRS}\sin(\varphi_l)cos(\vartheta_l)}, \dots, e^{-j kd_{IRS}(N_x-1)\sin(\varphi_l)cos(\vartheta_l)}],\\
   & \mathbf{b}_{l,z}= [1, e^{-j kd_{IRS}cos(\varphi_l)}, \dots, e^{-j kd_{IRS}(N_z-1)cos(\varphi_l)}],
\end{align}
where $d_{IRS}$ is the inter-element spacing,  and $\vartheta_l $ $(\varphi_l)$ denote the azimuth (elevation)  AoA  of the path from BS to IRS $l$  \cite{upachannel2}. 
{
It is apparent that each BS-IRS $l$ channel matrix has a rank one structure, thus $\mathbf{H}_{1,l}$ is a rank-one matrix with a unique non-zero singular value $\lambda_{\mathbf{H}_{1,l}} =  \sqrt{\beta_{1,l}}$  \cite{tsewirelesscomm}. The rank one assumption is accurate since the elements at both the BS and IRSs are co-located and the IRSs are assumed to be in the far-field, i.e.,  the propagation distance $d_{BS-IRS_l}$ between the BS and IRS $l$ is much larger than the largest dimension of that IRS. Although a channel with rank one means only one degree of freedom for the system, with distributed surfaces the overall channel between BS and each user will at least have rank $L$, which allows multiple users to be served simultaneously. 
}

 %To ensure far-field assumption holds true, we impose
%\begin{align}
%\label{eqfarfield}
    %d_{BS-IRS_l} \geq \frac{2D^2}{\lambda_c},
%\end{align}
%where $D$ is the largest dimension of the IRS which is the diagonal length of the UPA array.

\subsection{Problem Statement}

We focus on the average (ergodic) analysis of the SINR at each user in the distributed IRSs-assisted multi-user MISO system, where each IRS is associated with a user in the system. For a particular IRS-user association pair, we choose the design for IRS phase shifts that would maximize the received signal strength at that user. Under that design, we utilize statistical tools to obtain a closed form expression for the average SINR at each user. Therefore, we develop an average SINR expression under IRS-user association parameters, optimized IRS passive beamforming, and MRT precoding at the BS. Next, we formulate and solve a max-min average SINR optimization problem to find the optimal IRS-user association parameters.
  
\section{Downlink SINR Analysis }
\label{secavg}

With distributed IRSs deployed in a multi-user system, we consider associating IRSs to different users in an optimal manner to achieve a desired performance objective such as max-min average SINR in this work. We assume that each IRS will be associated with and therefore tuned to a single user. The motivation behind single-user association with an IRS is that it is user-centric and simplifies the optimal phase shifts design since the design can be based on the local CSI of the associated user, thus reducing complexity. A future research direction can be to consider the setting where each IRS is associated to multiple users and is optimally tuned to serve those users. Using  \eqref{eqtransmit}, the $k^{th}$ user receives %The results we obtain with the assumption perfect CSI are relevant since they pave the way for imperfect CSI analysis and can serve as meaningful upper bounds for imperfect CSI scenarios.% the BS sends the transmit signal vector $  \mathbf{x}$ as
%\begin{align}
 %   \mathbf{x}=\sum_{k=1}^K\sqrt{p_k}\mathbf{f}_ks_k.
%\end{align}
%The transmit power for user $k$ is $p_k$, while $\mathbf{f}_k \in \mathbb{C}^{M \times 1}$ is a precoding vector representing the spatial direction of the signal. Furthermore, $s_k \sim \mathcal{CN}(0,1)$ is the intended Gaussian signal for user $k$ with normalized unit power. %Since the BS has an antenna array, the $\mathbf{f}_k$ is also called the precoding vector which is often chosen to be spatially directive to a particular user. The number of antenna elements $M$ in the array has a relation to the beam-width of the radiated signal from the array. Larger $M$ leads to a smaller beam-width and a focused beam which will make the radiated signal directive for a user and lessen the interference to other users as intuitively it will occupy less space. One takeway is that larger $M$ is better for SINRs at the system level. 
\begin{align}
\label{Eqrec1}
y_{k}&=\mathbf{h}_{d,k}^H\mathbf{x}+\sum_{l=1}^{L}\mathbf{h}_{2,l,k}^H \boldsymbol{\Theta}_l^H \mathbf{H}_{1,l}^H\mathbf{x}+n_{k}, 
\end{align}
where the channels are given in \eqref{model_h2lk}, \eqref{model_hdk}, and \eqref{eqmodel_H1}. To aid in expressing the instantaneous SINR $\gamma_k$ for user $k$, we rewrite $y_{k}$ as $y_{k}={y_k}_D+{y_k}_{I}$, where ${y_k}_D$ is the desired signal given by
\begin{align}
{y_k}_D&=\sqrt{p}_k\mathbf{h}_{d,k}^H\mathbf{f}_ks_k+\sqrt{p}_k\sum_{l=1}^{L}\mathbf{h}_{2,l,k}^H \boldsymbol{\Theta}_l^H \mathbf{H}_{1,l}^H\mathbf{f}_ks_k,
\end{align}
and ${y_k}_{I}$ is the interference and noise terms given by
\begin{align}
{y_k}_{I}&= \sum_{t\neq k, t=1}^{K} \bigg(\sqrt{p}_t\mathbf{h}_{d,k}^H\mathbf{f}_t s_t+\sqrt{p}_t\sum_{l=1}^{L}\mathbf{h}_{2,l,k}^H \boldsymbol{\Theta}_l^H \mathbf{H}_{1,l}^H\mathbf{f}_ts_t\bigg)+n_k, 
\end{align}

Thus, we construct the SINR $\gamma_k$ at user $k$ under the transmission model just described as
\begin{align}
\label{eqinstSINR1}
        \gamma_{k}=\frac{p_k\|\mathbf{h}^H_{k} \mathbf{f}_{k}\|^2}{\sum_{t\neq k, t=1}^{K}p_t\|\mathbf{h}^H_{k} \mathbf{f}_{t}\|^2+\sigma^2}.
\end{align}

The direct and IRS cascaded channels in the downlink can be rewritten collectively as
\begin{align}\label{eqcollect}
    \mathbf{h}^H_{k}&=\mathbf{h}_{d,k}^H+\sum_{l=1}^{L}\mathbf{h}_{2,l,k}^H \boldsymbol{\Theta}_l^H \mathbf{H}_{1,l}^H=\mathbf{h}_{d,k}^H+\sum_{l=1}^{L}\mathbf{v}_{l}^H  \mathbf{H}_{0,l,k}^H,
\end{align}
where 
\begin{align}
\label{eqholk}
\mathbf{H}_{0,l,k}=\mathbf{H}_{1,l}\text{diag}(\mathbf{h}_{2,l,k})= \sqrt{\beta_{1,l}}\mathbf{a}_l\mathbf{b}_l^H \text{diag}(\mathbf{h}_{2,l,k})\end{align}
 and $\mathbf{v}_l=\text{diag}(\boldsymbol{\Theta}_l) \in \mathbb{C}^{N \times 1}$.
 Using \eqref{eqcollect}, we rewrite the instantaneous SINR $\gamma_k$ in \eqref{eqinstSINR1} as 
 \begin{align}
\label{eqinstSINRexp2}
        \gamma_{k}=\frac{p_k\|(\mathbf{h}_{d,k}^H+\sum_{l=1}^{L}\mathbf{v}_{l}^{H} \mathbf{H}_{0,l,k}^H ) \mathbf{f}_{k}\|^2}{\sum_{t\neq k, t=1}^{K}p_t\|(\mathbf{h}_{d,k}^H+\sum_{l=1}^{L}\mathbf{v}_{l}^{H}  \mathbf{H}_{0,l,k}^H)\mathbf{f}_{t}\|^2+\sigma^2}.
\end{align}
At this stage, we introduce the association variables $\lambda_{l,k} \in \{0,1\}$ which are binary variables denoting association between the $l^{th}$ IRS and $k^{th }$ user. Note that $\lambda_{l,k} =1$ when the $l^{th}$ IRS is associated to the $k^{th}$ user, and  $\lambda_{l,k}=0$ when there is no association with that user. The overall channel in \eqref{eqcollect} with association variables can be reformulated as 
\begin{align}
\label{eqvlambda}
  \mathbf{h}^H_{k}(    \boldsymbol{\lambda}_{k})=\mathbf{h}_{d,k}^H+\sum_{l=1}^{L}\lambda_{l,k}\mathbf{v}_{l}^{{k_l}^H} \mathbf{H}_{0,l,k}^H+\sum_{l=1}^{L}(1-\lambda_{l,k})\mathbf{v}_{l}^{k_l^H}  \mathbf{H}_{0,l,k}^H, 
  \end{align}
where $\boldsymbol{\lambda}_{k} \in \mathbb{B}^{1 \times L}$ denotes the array of associated and/or non-associated IRSs with respect to user $k$ and is a row vector in $\boldsymbol{\Lambda\in \mathbb{B}^{K \times L}}$  which is the IRS-user association matrix. To make the association between the optimized (tuned) beamforming vector $\mathbf{v}_{l}$ to a particular user $k_l$, we replace $\mathbf{v}_{l}$ by $\mathbf{v}_{l}^{k_l}$ shown in \eqref{eqirsphase} henceforth, where $k_l$ is the user for which $\lambda_{l,k_l}=1$. For \eqref{eqvlambda}, $k_l=k$ in the second term $\sum_{l=1}^{L}\lambda_{l,k}\mathbf{v}_{l}^{{k_l}^H} \mathbf{H}_{0,l,k}^H$ which means $\lambda_{l,k}=1$. As for the third term $\sum_{l=1}^{L}(1-\lambda_{l,k})\mathbf{v}_{l}^{k_l^H}  \mathbf{H}_{0,l,k}^H$ we have $\mathbf{v}_{l}^{k_l}$ is the beamforming vector $\mathbf{v}_{l}$ of IRS $l$  optimized for associated user $k_l$ where $k_l \neq k$  such that $\lambda_{l,k_l}=1$ and  $\lambda_{l,k}=0$. The notation $k_l$ is to ascertain an associated user $k_l$ which depends on its IRS association. In addition, we can rewrite \eqref{eqtransmit} with association parameters as $
y_{k}= \mathbf{h}^H_{k}(\boldsymbol{\lambda}_{k})\mathbf{x}+n_{k}, $
and the updated, instantaneous SINR as 
 \begin{align}
\label{eqinstSINR2}
        \gamma_{k}=\frac{p_k\|\mathbf{h}^H_{k}(\boldsymbol{\lambda}_{k}) \mathbf{f}_{k}\|^2}{\sum_{t\neq k, t=1}^{K}p_t\|\mathbf{h}^H_{k}(\boldsymbol{\lambda}_{k})\mathbf{f}_{t}\|^2+\sigma^2}.
\end{align}
An important design goal is to optimize the downlink SINR in \eqref{eqinstSINR2} with respect to the IRSs configuration, which we split into IRSs passive beamforming and IRS-user association. In the next section, we find the optimal reflect beamforming vector $\mathbf{v}_{l}^{k_l} $ at each IRS $l$ that would enhance the transmission quality to its associated user $k_l$.

\subsection{IRS Passive Beamforming}
We find the optimal passive beamforming vector $\mathbf{v}_{l}^{k_l} $ for IRS $l$ associated with user $k_l$ such that $\lambda_{l,k_l}=1$, so as to maximize the channel gain via that IRS to that user.  Note that finding $\mathbf{v}_{l}^{k_l} $ for each IRS $l$ such that SINR at its associated user $k_l$ is maximized will result in a highly intractable joint optimization problem involving the beamforming vectors of all IRSs and their association parameters with the users. There is no known closed-form optimal solution for $\mathbf{v}_{l}$ that maximizes the SINR, which is why we resort to maximizing the channel gain instead, similar to \cite{IRSuserassociation}. Hence, the passive beamforming optimization problem for an IRS $l$ associated with user $k_l$ can be defined as
\begin{align}
   (P0) \hspace{0.2in}\underset{\mathbf{v}_l}{\text{max}}&\hspace{0.2in}\|\mathbf{h}_{d,k_l}^H+\mathbf{v}_{l}^{H}  \mathbf{H}_{0,l,k_l}^H\|^2 \\ 
   &\text{s.t.} \hspace{0.2in}|v_{l,n}|=1. \forall n.
\end{align}
Here, $v_{l,n}$ denotes the $n^{th}$ element in $\mathbf{v}_l$, $\mathbf{h}_{d,k_l}$ and $\mathbf{H}_{0,l,k_l}$ are the direct and IRS cascaded channels for user $k_l$. Expanding the objective function yields $\|\mathbf{h}_{d,k_l}^H\|^2+2\langle\mathbf{v}_{l} ,\mathbf{H}_{0,l,k_l}^H\mathbf{h}_{d,k_l} \rangle+\|\mathbf{v}_{l}^H  \mathbf{H}_{0,l,k_l}^H\|^2$, where we can drop the first term since it does not depend on the optimization variable to get
\begin{align}
\label{eqp0optimal}
   (P0') \hspace{0.2in}\underset{\mathbf{v}_l}{\text{max}}\hspace{0.2in}& 2\langle\mathbf{v}_{l} ,\mathbf{H}_{0,l,k_l}^H\mathbf{h}_{d,k_l} \rangle+\|\mathbf{v}_{l}^{H} \mathbf{H}_{0,l,k_l}^H\|^2\\ 
   \label{constraintvl}
   &\text{s.t.} \hspace{0.2in}|v_{l,n}|=1. \forall n.
\end{align}

%Note that $P0$ finds the optimized beamforming vector $\mathbf{v}_l^{k}$ for user $k$, when $\lambda_{l,k}=1$. 

\begin{lemma}The  optimal solution of $(P0')$ for the beamforming vector of IRS $l$ associated with user $k_l$, such that $\lambda_{l,k_l}=1$ is
\begin{align}
\label{eqirsphase}
  \mathbf{v}_{l}^{k_l}=e^{j\angle{\text{diag}(\mathbf{h}^H_{2,l,k_l})\mathbf{b}_l}}e^{j\angle{\mathbf{a}_l^H\mathbf{h}_{d,k_l}}}.
\end{align}

\end{lemma}
\begin{IEEEproof}
%To find the angle of a complex (Hermitian) inner product, %Where $\Re{(\langle\cdot,\cdot\rangle)}$ denotes the real part of the inner product of two complex vectors,
It suffices to express  $\langle\mathbf{v}_{l} ,\mathbf{H}_{0,l,k_l}^H\mathbf{h}_{d,k_l} \rangle= \sum_{n=1}^{N}|v_{l,n}||{h_m}|\cos(\angle v_{l,n}-\angle h_m)$, where $v_{l,n}$ (${h}_m$) is the $n^{th}$ component of $\mathbf{v}_{l}$ ($\mathbf{H}_{0,l,k_l}^H\mathbf{h}_{d,k_l}$). This expression achieves its maximum value when $\angle v_{l,n}^{k_l}=\angle{h}_{m}$, equivalently ($\angle\mathbf{v}_{l}^{k_l} =\angle  \text{diag}(\mathbf{h}_{2,l,k_l}^H)\mathbf{b}_l\mathbf{a}_l^H\mathbf{h}_{d,k_l})$ where $v_{l,n}^{k_l}$ is the $n^{th}$ element of $\angle\mathbf{v}_{l}^{k_l}$, which is solution presented in \eqref{eqirsphase} in  simplified form.
Similar argument can be done for the second term in the objective function \eqref{eqp0optimal}, where
  $\underset{\mathbf{v}_{l}}{\text{max}}\hspace{0.2in} \|\mathbf{v}_l^H  \mathbf{H}_{0,l,k_l}^H\|^2$
  \begin{align}
 &=\underset{\mathbf{v}_l}{\text{max}}\hspace{0.2in}\mathbf{v}_l^H \text{diag}(\mathbf{h}_{2,l,k_l}^H)\mathbf{b}\mathbf{a}^H\mathbf{a}\mathbf{b}^H \text{diag}(\mathbf{h}_{2,l,k_l})\mathbf{v}_l =\underset{\mathbf{v}_l}{\text{max}}\hspace{0.2in}\|\mathbf{a}_l\|^2|\mathbf{v}_l^H \text{diag}(\mathbf{h}_{2,l,k_l}^H)\mathbf{b}_l|^2,
    \\
    \label{eqvaryingv}
    &=\|\mathbf{a}_l\|^2\bigg|\sum_{n=1}^N|{v}^k_{l,n}||{h}_{2,l,k_l,n}||{b}_{l,n}|e^{j({\angle{h_{2,l,k_l,n}}+\angle{b^*_{l,n}}+\angle{h^*_{2,l,k_l,n}}+\angle{b_{l,n}}}+\angle{\mathbf{h}_{d,k_l}^H\mathbf{a}_l})}\bigg|^2,
      \\
    \label{eqvaryingv2}
    &=\|\mathbf{a}_l\|^2(\sum_{n=1}^N|{h}_{2,l,k_l,n}||{b}_{l,n}|)^2|e^{j(\angle{\mathbf{h}_{d,k_l}^H\mathbf{a}_l})}|^2
   % \label{eqmaxvalue}
    =\|\mathbf{a}_l\|^2(\sum_{n=1}^N|{h}_{2,l,k_l,n}||{b}_{l,n}|)^2.
\end{align}
Here,  \eqref{eqvaryingv2} is obtained by  noting that there is a unit-modulus constraint on $v_{l,n}$, i.e., $|v_{l,n}|=1$. Moreover, the last line is found by recalling that the absolute square of a complex number is the complex number multiplied by its conjugate, i.e.,  $|e^{j(\angle{\mathbf{h}_{d,k_l}^H\mathbf{a}_l})}|^2=e^{j(\angle{\mathbf{h}_{d,k_l}^H\mathbf{a}_l})}e^{-j(\angle{\mathbf{h}_{d,k_l}^H\mathbf{a}_l})}=e^{j(\angle{\mathbf{h}_{d,k_l}^H\mathbf{a}_l}-\angle{\mathbf{h}_{d,k_l}^H\mathbf{a}_l})}=1$. Finally, the expression achieves its maximum value when the argument of all the complex numbers involved cancel each other which is accomplished with the choice of reflection beamforming vector given in \eqref{eqirsphase}. 
\end{IEEEproof}

Next, we turn our attention to obtaining the closed-form expression for the average SINR at each user in \eqref{eqinstSINR2} under the IRS  passive beamforming design in \eqref{eqirsphase}. 

\subsection{Average SINR}
\label{secirsuserassoc}
    Since practical wireless channels undergo random fading, an important performance metric is the average SINR $\bar{\gamma}_k$ which depends only on statistics of the channels. Channel statistics such as path loss and correlation matrices change slowly as compared to the fast fading channels themselves. Therefore, instead of computing instantaneous SINR in each coherence interval, $\bar{\gamma}_k$ can be used as a performance metric to optimize the system. In fact, by using $\bar{\gamma}_k$ as a performance metric, the IRS-user association parameters will not need to be updated on the coherence time scale but only when the large-scale statistics of the channels change. Once the association is established, the BS will use local CSI of the associated user to determine each IRS's optimal configuration in \eqref{eqirsphase}. Next, we present the approximation of the average SINR in Lemma \ref{lemavg}.

\begin{lemma}
\label{lemavg}
The average SINR $\bar{\gamma}_k= \mathbb{E}[\gamma_{k}]$, with $\gamma_{k}$ given by \eqref{eqinstSINR2} is approximated as \cite{sinr_Ergodic}, \cite{sinr_Taylor1}
\begin{align}
\label{eqsinravg1}
       \bar{\gamma}_k= \mathbb{E}[\gamma_{k}]\approx \frac{p_k \mathbb{E}[\|\mathbf{h}^H_{k}(\boldsymbol{\lambda}_{k}) \mathbf{f}_{k}(\boldsymbol{\lambda}_{k})\|^2]}{\sum_{t\neq k, t=1}^{K}p_t\mathbb{E}[\|\mathbf{h}^H_{k}(\boldsymbol{\lambda}_{k}) \mathbf{f}_{t}(\boldsymbol{\lambda}_{t})\|^2]+\sigma^2}
\end{align}
\end{lemma}
\begin{IEEEproof}
Consider the expectation $\mathbb{E}[\frac{X}{Y}]$ and assume $\mathbb{E}[X]=\mu_X$ and $\mathbb{E}[Y]=\mu_Y$. We can expand the ratio $\frac{X}{Y}$ (using the bivariate first-order Taylor series expansion) around the point $(\mu_X,\mu_Y)$ so that 
\begin{align}
        \frac{X}{Y}=\frac{\mu_X}{\mu_Y}+\frac{1}{\mu_Y}(X-\mu_X)-\frac{\mu_X}{\mu_Y^2}(Y-\mu_Y)+C
\end{align}
where $C$ denotes the remaining higher-order terms in the expansion. Taking the first moment of this ratio provides 
\begin{align}
        &\mathbb{E}\left[\frac{X}{Y}\right]= \mathbb{E}\left[\frac{\mu_X}{\mu_Y}+\frac{1}{\mu_Y}(X-\mu_X)-\frac{\mu_X}{\mu_Y^2}(Y-\mu_Y)+C\right] \\ &\underset{(i)}{\approx} \mathbb{E}\left[\frac{\mu_X}{\mu_Y}\right]+\mathbb{E}\left[\frac{1}{\mu_Y}(X-\mu_X)\right]-\mathbb{E}\left[\frac{\mu_X}{\mu_Y^2}(Y-\mu_Y)\right]\underset{(ii)}{=}  \frac{\mathbb{E}[X]}{\mathbb{E}[Y]}
\end{align}
where $(i)$ follows from dropping the higher order terms $C$, and $(ii)$ follows since the expectation of the second and third terms is zero. Applying this to $\mathbb{E}[\gamma_{k}]$ yields  \eqref{eqsinravg1}.
\end{IEEEproof}

%\textit{Remark 1} This approximation for $ \mathbb{E}[\gamma_{k}]$ is justified as follows: since the numerator and denominator of $\bar{\gamma_k}$ contain quadratic forms, their variance relative to the mean is small  \cite{sinr_Ergodic}, \cite{sinr_Taylor1}, \cite{SINR_TAYLOR2}. 

This approximation of average SINR is very tight as will be  verified numerically in Fig. \ref{figtight} in the simulation results. 
Substituting the MRT precoder \eqref{eqfk} in \eqref{eqsinravg1} yields
\begin{align}
\label{equndermRTwithassocHARD}
        \bar{\gamma}_{k}=\frac{p_k \frac{\mathbb{E}[\|\mathbf{h}^H_{k}(\boldsymbol{\lambda}_{k})\|^4]}{\mathbb{E}[\|\mathbf{h}^H_{k}(\boldsymbol{\lambda}_{k})\|^2]}}{\sum_{t\neq k, t=1}^{K}p_t\frac{\mathbb{E}[\|\mathbf{h}^H_{k}(\boldsymbol{\lambda}_{k}) \mathbf{h}_{t}(\boldsymbol{\lambda}_{t})\|^2]}{\mathbb{E}[\|\mathbf{h}^H_{t}(\boldsymbol{\lambda}_{t})\|^2]}+\sigma^2}
\end{align}

\subsection{Main Results}
\label{secmainresults}

Before we present the main derivation for the average SINR, note that the channel $\mathbf{h}_{k}(\boldsymbol{\lambda}_{k})$ in \eqref{eqvlambda} is distributed as $\mathbf{h}_{k}(\boldsymbol{\lambda}_{k})  \sim \mathcal{CN}(0, \mathbf{R}_{k})$ since it contains the addition of complex Gaussian vectors. As a result, $\mathbf{h}_{k}(\boldsymbol{\lambda}_{k})$ is a complex Gaussian vector with zero mean and correlation matrix $\mathbf{R}_{k}$, which will be derived later in this section.

In this section, we derive all the expectations in \eqref{equndermRTwithassocHARD}, which would require us to find the first and second moments of a complex Gaussian quadratic form abbreviated as (CGQF). First, we expand the expectation of the squared channel gain as
\begin{align}
\label{eqnormsqexpand}
&\mathbb{E}[\|{\mathbf{h}}_{k}(\boldsymbol{\lambda}_{k})\|^2]=\mathbb{E}[\mathbf{h}_{d,k}^H\mathbf{h}_{d,k}+2\sum_{l=1}^{L}\lambda_{l,k}\mathbf{v}_{l}^{{k_l}^H}  \mathbf{H}_{0,l,k}^H\mathbf{h}_{d,k}+\\  &\nonumber\sum_{l=1}^{L}\sum_{\bar{l}=1}^{L}\lambda_{l,k}\lambda_{\bar{l},k}\mathbf{v}_{l}^{{k_l}^H}  \mathbf{H}_{0,l,k}^H\mathbf{H}_{0,\bar{l},k}\mathbf{v}_{\bar{l}}^{{k_{\bar{l}}}}+\sum_{l=1}^{L}\sum_{ \nonumber\bar{l}=1}^{L}(1-\lambda_{l,k})(1-\lambda_{\bar{l},k})\mathbf{v}_{l}^{k_l^H}  \mathbf{H}_{0,l,k}^H\mathbf{H}_{0,\bar{l},k}\mathbf{v}_{\bar{l}}^{{k_{\bar{l}}}}\\ \nonumber &+2\sum_{l=1}^{L}(1-\lambda_{l,k})\mathbf{v}_{l}^{k_l^H}  \mathbf{H}_{0,l,k}^H\mathbf{h}_{d,k}+2\sum_{l=1}^{L}\sum_{\bar{l}=1}^{L}(1-\lambda_{\bar{l},,k})\lambda_{{l,k}}\mathbf{v}_{l}^{{k_l}^H}  \mathbf{H}_{0,l,k}^H\mathbf{H}_{0,\bar{l},k}\mathbf{v}_{\bar{l}}^{{k_{\bar{l}}}}],
    \end{align}
 where $\mathbf{v}^{k_l}_l$ and $\mathbf{v}^{k_{\bar{l}}}_{\bar{l}}$ are of the form in \eqref{eqirsphase} for user $k_l$ associated with IRS $l$ and user $k_{\bar{l}}$ associated with IRS $\bar{l}$, respectively. Note that the last two terms in \eqref{eqnormsqexpand} are zero due to independence between channels $\mathbf{h}_{2,l,k}$'s and $\mathbf{h}_{d,k}$'s. Next we present this expectation in a closed form in the following lemma.

    \begin{lemma}
    \label{lemmanormhk2}
The expectation in \eqref{eqnormsqexpand} is given by
        \begin{align}
        \label{eqlemmanormeq}
 \mathbb{E}[\|{\mathbf{h}}_{k}(\boldsymbol{\lambda}_{k})\|^2]=&M\beta_{d,k}+\sum_{l=1}^{L}\lambda_{l,k}\bigg(\alpha+tr(\mathbf{H}_{1,l}^H\mathbf{H}_{1,{l}}\boldsymbol{\Sigma}_{\tilde{\mathbf{v}}_{l}^{k_l}}-\beta_{2,l,k}\mathbf{H}_{1,l}^H\mathbf{H}_{1,{l}}) \bigg)\nonumber \\
&+\beta_{2,l,k}tr(\mathbf{H}_{1,l}^H\mathbf{H}_{1,{l}}),
    \end{align}
    \end{lemma}
where $\alpha=\sqrt{\beta_{1,l}\beta_{2,l,k}\beta_{d,k}}\frac{\pi\sqrt{M}{N}}{2}$ and  $\boldsymbol{\Sigma}_{\tilde{\mathbf{v}}_{l}^{k_l}}=
        \beta_{2,l,k}\mathbf{I}_N+
        \frac{\pi\beta_{2,l,k}}{4}
       \mathbf{e}^{j\angle\mathbf{H}_{1,l}^H\mathbf{H}_{1,l}}\odot(\mathbf{1}_N-\mathbf{I}_N)$.
 \begin{IEEEproof}
 The proof is postponed to Appendix \ref{appendixIEEEproofchannelgainsq}.
 \end{IEEEproof}

Note that       $\mathbb{E}[\|{\mathbf{h}}_{k}(\lambda_{l,k})\|^2]=tr(\mathbf{R}_{k})$. The term in $ \mathbb{E}[\|{\mathbf{h}}_{k}(\boldsymbol{\lambda}_{k})\|^2]$ accounting for the gains from  non-associated IRSs is $\sum_{l=1}^{L}(1-\lambda_{l,k})\beta_{2,l,k}tr(\mathbf{H}_{1,l}^H\mathbf{H}_{1,{l}})$ which has been distributed for a more concise representation in \eqref{eqlemmanormeq}.  It can be seen that the average squared channel gain increases with respect to $M$. On the other hand, the average squared channel gain depends on the user association parameters as well as  $L$ and $N$. For instance, if user $k$ does not have any IRSs associated to it, i.e. $\lambda_{l,k}=0, \forall l$, then we still gain from all the non-associated IRSs although they are optimized for other users.
Next we present in Lemma \ref{lemma4th} the expression of $\mathbb{E}[\|{\mathbf{h}}_{k}(\lambda_{l,k})\|^4]$, which  will be used to complete the analysis of the numerator of $\bar{\gamma}_k$.
 
           \begin{lemma}   
           \label{lemma4th}
 The fourth moment $\mathbb{E}[\|{\mathbf{h}}_{k}(\lambda_{l,k})\|^4]$ in the numerator of $\bar{\gamma}_k$ in \eqref{equndermRTwithassocHARD} is derived using a result on the second moment of CGQF  as 
              \begin{align}
         &\mathbb{E}[\|{\mathbf{h}}_{k}(\boldsymbol{\lambda}_{k})\|^4]=
       tr(\mathbf{R}_k^2)+tr(\mathbf{R}_{k})^2. \label{fourthmom1}
           \end{align}
        \end{lemma}
        \begin{IEEEproof}
          \begin{align}
        &\mathbb{E}[\|{\mathbf{h}}_{k}(\boldsymbol{\lambda}_{k})\|^4]= \text{Var}[\|{\mathbf{h}}_{k}(\boldsymbol{\lambda}_{k})\|^2]+(\mathbb{E}[\|{\mathbf{h}}_{k}(\boldsymbol{\lambda}_{k})\|^2])^2\\ \label{fourthmom2}
        &=\text{Var}[\|{\mathbf{h}}_{k}(\boldsymbol{\lambda}_{k})\|^2]+tr(\mathbf{R}_{k})^2  =tr(\mathbf{R}_k^2)+tr(\mathbf{R}_{k})^2,
           \end{align}
           where $\text{Var}[\|{\mathbf{h}}_{k}(\boldsymbol{\lambda}_{k})\|^2]=tr(\mathbf{R}_k^2)$ follows from the second moment of a CGQF \cite{complexGCF}, \cite{sinr_Ergodic}.
         
        \end{IEEEproof}
      
{ One can see from Lemma \ref{lemma4th} that this result relies heavily on the correlation matrix $\mathbf{R}_k$, as its diagonal elements become larger so does the channel gain. In fact, increasing $M$ increases the fourth-order moment in \eqref{fourthmom1} quadratically. However, increasing the size of the diagonal elements of $\mathbf{R}_k$ also affects the interference as seen below. }
       
  Next we present in Lemma \ref{lemmaint} the expression of $\mathbb{E}[\|\mathbf{h}_{k}^H(\boldsymbol{\lambda}_{k})\mathbf{h}_{t}(\boldsymbol{\lambda}_{t})\|^2]$ which appears in the interference term in \eqref{equndermRTwithassocHARD}.
	
        \begin{lemma}
        \label{lemmaint}
In the interference term $\sum_{t\neq k, t=1}^{K}p_t\frac{\mathbb{E}[\|\mathbf{h}^H_{k}(\boldsymbol{\lambda}_{k}) \mathbf{h}_{t}(\boldsymbol{\lambda}_{t})\|^2]}{\mathbb{E}[\|\mathbf{h}^H_{t}(\boldsymbol{\lambda}_{t})\|^2]}$, the numerator is found to be
        \begin{align}
        \label{eqpostrrkrt}
        &\mathbb{E}[\|\mathbf{h}_{k}^H(\boldsymbol{\lambda}_{k})\mathbf{h}_{t}(\boldsymbol{\lambda}_{t})\|^2]=tr(\mathbf{R}_{t}\mathbf{R}_{k}).
           \end{align}
           
        \end{lemma}
        \begin{IEEEproof}    \begin{align}
          &\mathbb{E}[\|\mathbf{h}_{k}^H(\boldsymbol{\lambda}_{k})\mathbf{h}_{t}(\boldsymbol{\lambda}_{t})\|^2]=
           \mathbb{E}[\mathbf{h}_{k}^H(\boldsymbol{\lambda}_{k})\mathbf{h}_{t}(\boldsymbol{\lambda}_{t})\mathbf{h}_{t}^H(\boldsymbol{\lambda}_{t})\mathbf{h}_{k}(\boldsymbol{\lambda}_{k})].
              \end{align}
            Since $\mathbf{h}_{t}(\boldsymbol{\lambda}_{t})$ and $\mathbf{h}_{k}(\boldsymbol{\lambda}_{k})$ are independent, we can use conditional expectation as
           \begin{align}     
        &\mathbb{E}_{\mathbf{h}_{k}}[\mathbf{h}_{k}^H(\lambda_{l,k})\mathbb{E}_{\mathbf{h}_{t}}[\mathbf{h}_{t}(\lambda_{l,k})\mathbf{h}_{t}^H(\lambda_{l,k})|\mathbf{h}_{k}(\lambda_{l,k})]\mathbf{h}_{k}(\lambda_{l,k})] \label{eqrkrt}=\mathbb{E}_{\mathbf{h}_{k}}[\mathbf{h}_{k}^H(\lambda_{l,k})\mathbf{R}_{t}\mathbf{h}_{k}(\lambda_{l,k})],\\ &=\mathbb{E}_{\mathbf{h}_{k}}[tr(\mathbf{h}_{k}^H(\lambda_{l,k})\mathbf{R}_{t}\mathbf{h}_{k}(\lambda_{l,k}))]=\mathbb{E}_{\mathbf{h}_{k}}[tr(\mathbf{R}_{t}\mathbf{h}_{k}(\lambda_{l,k})\mathbf{h}^H_{k}(\lambda_{l,k}))],\\\label{eqtrrtrk}&=tr(\mathbf{R}_{t}\mathbb{E}_{\mathbf{h}_{k}}[\mathbf{h}_{k}(\lambda_{l,k})\mathbf{h}^H_{k}(\lambda_{l,k})]) 
        =tr(\mathbf{R}_{t}\mathbf{R}_{k}).
           \end{align} 
                   The denominator of the interference term is derived in Lemma \ref{lemmanormhk2}, previously.
        \end{IEEEproof}
% It is straightforward to note that when $tr(\mathbf{R}_k)$ is large, the overall interference is augmented for other users. Intuitively, increasing the channel gain for one user, increases the interference sensed at another.
To obtain expressions in   \eqref{fourthmom1} and \eqref{eqpostrrkrt} we need to find the expression for $   \mathbf{R}_{k}$ which is next computed. Define the correlation matrix for $\mathbf{h}_k$ to be
           \begin{align}
           \label{eqrkexp}
        \mathbf{R}_{k}=&\mathbb{E}[{\mathbf{h}}_{k}(\boldsymbol{\lambda}_{k}){\mathbf{h}}_{k}(\boldsymbol{\lambda}_{k})^H]=\mathbb{E}[\mathbf{h}_{d,k}\mathbf{h}_{d,k}^H+2\sum_{l=1}^{L}\lambda_{l,k}\mathbf{h}_{d,k} \mathbf{v}_{l}^{{k_l}^H} \mathbf{H}_{0,l,k}^H\\ \nonumber &+\sum_{l=1}^{L}\sum_{\bar{l}=1}^{L}\lambda_{l,k}\lambda_{\bar{l},k} \mathbf{H}_{0,l,k}\mathbf{v}_{l}^{k_l} \mathbf{v}_{\bar{l}}^{{k_{\bar{l}}}^H} \mathbf{H}_{0,\bar{l},k}^H+\sum_{l=1}^{L}\sum_{\bar{l}=1}^{L}(1-\lambda_{l,k})(1-\lambda_{\bar{l},k})  \mathbf{H}_{0,l,k}\mathbf{v}_{l}^{k_l}\mathbf{v}_{\bar{l}}^{k_{\bar{l}}^H}\mathbf{H}_{0,\bar{l},k}^H],
           \end{align}
        where other terms are zeros as shown in Appendix \ref{appendixIEEEproofchannelgainsq}.
        The expression given in the next lemma for $\mathbf{R}_k$ depends on the system dimensions $M, N, L$, path loss factors $\beta_{1,l}, \beta_{2,l,k},\beta_{d,k} $, and association parameters $\lambda_{l,k}$s. In Sec. \ref{secirsuserassoc}, we optimize the association parameter matrix $\boldsymbol{\Lambda}$ as to maximize the minimum $\bar{\gamma}_{k}$.
        \begin{lemma}
      \label{lemmaRk}
The correlation matrix for channel $\mathbf{h}_k(\boldsymbol{\lambda}_k)$ in \eqref{eqvlambda} is shown to be
        \begin{align}
        \label{eqfinalrk}
     \mathbf{R}_{k}=&\beta_{d,k}\mathbf{I}_M+\sum_{l=1}^{L}\lambda_{l,k}\bigg(2\sqrt{\beta_{1,l}\beta_{2,l,k}\beta_{d,k}}\frac{N\pi}{4\sqrt{M}} \mathbf{e}^{j\angle\mathbf{H}_{1,l}\mathbf{H}_{1,l}^H}\\ \nonumber &+ \mathbf{H}_{1,l}
       \boldsymbol{\Sigma}_{\tilde{\mathbf{v}}_{l}^{k_l}} \mathbf{H}_{1,l}^H- \beta_{2,l,k}\mathbf{H}_{1,l}\mathbf{H}_{1,{l}}^H\bigg)+\sum_{l=1}^{L}\beta_{2,l,k}\mathbf{H}_{1,l}\mathbf{H}_{1,{l}}^H.
           \end{align}  
             
        \end{lemma}  
        \begin{IEEEproof}
        The proof is postponed to Appendix \ref{AppendDerivationRK}.
        \end{IEEEproof}
       {Now, we can combine these results in the following Theorem \ref{theoremsinravg} to find a closed-form expression for the average SINR at user $k$. 
}   \begin{theorem}
           \label{theoremsinravg}
Using the results from Lemmas \ref{lemmanormhk2}, \ref{lemma4th}, \ref{lemmaint}, the ergodic SINR under MRT and IRS passive beamforming in \eqref{eqirsphase} for a given IRS-user association matrix $\boldsymbol{\Lambda}$ is given as
        \begin{align}
\label{eqthmsimpletogetheravgSINR}
      \bar{\gamma}_{k}=\frac{ \frac{p_k}{tr(\mathbf{R}_{k})}\big(tr(\mathbf{R}_k^2)+tr(\mathbf{R}_{k})^2\big)}{\sum_{t\neq k, t=1}^{K}\frac{p_t}{tr(\mathbf{R}_{t})}tr(\mathbf{R}_{t}\mathbf{R}_{k})+\sigma^2}=\frac{ {c_k}(tr(\mathbf{R}_k^2)+tr(\mathbf{R}_{k})^2)}{\sum_{t\neq k, t=1}^{K}{c_t}tr(\mathbf{R}_{t}\mathbf{R}_{k})+\sigma^2},
\end{align}
           \end{theorem}
    where $c_k$ satisfies ${p_k}=c_k tr(\mathbf{R}_{k}) \forall k$, and  $\mathbf{R}_{k}$ is defined in \eqref{eqfinalrk} as a function of $\boldsymbol{\Lambda}$.

\begin{corollary}
\label{corortho}
When the correlation matrices of different users are orthogonal, i.e. $\mathbf{R}_{t}\mathbf{R}_{k}=\mathbf{0}_{M}$, then the average SINR in \eqref{eqthmsimpletogetheravgSINR} simplifies to an average SINR upper bound given by
 \begin{align}
\label{simplerSINRup}
         \bar{\gamma}_{k,Up}=\frac{{c_k}(tr(\mathbf{R}_k^2)+tr(\mathbf{R}_{k})^2)}{\sigma^2}
\end{align}
When the correlation matrices of all users are identical and the allocated powers for all users are the same, i.e., $\mathbf{R}_{t}=\mathbf{R}_{k}=\mathbb{E}[{\mathbf{h}}_{k}(\boldsymbol{\lambda}_{k}){\mathbf{h}}_{k}(\boldsymbol{\lambda}_{k})^H]$ \cite{massivemimobook}, and $p_1=p_2=\dots=p_K$, the average SINR simplifies to the following average SINR lower bound
\begin{align}
\label{simplerSINRlow}
         \bar{\gamma}_{k,Low}=\frac{ 1+tr(\mathbf{R}_{k})^2/tr(\mathbf{R}_{k}^2)}{(K-1)+\sigma^2/c_k tr(\mathbf{R}_{k}^2)}.
\end{align}

\end{corollary}

Essentially, Corollary \ref{corortho} presents meaningful bounds $  \bar{\gamma}_{k,Low} \leq \bar{\gamma}_{k} \leq   \bar{\gamma}_{k,Up} $ which are illustrated in Fig. \ref{figupperlower} in the simulations in Sec. \ref{secoptimizedIRS}, that show how the level of diversity between correlation matrices of different users impacts the average SINR. Having diverse channels lends to a higher overall SINR, while having users with similar correlation matrices  leads to a degradation in SINR performance. In the next section, we formulate the max-min SINR problem using the average SINR derived in \eqref{eqthmsimpletogetheravgSINR} to find the optimal IRS-user association pairs.
%which we can take an upper bound of the first term in the equality, and show that it is less which is sufficient.
%\begin{align}
    %&\sum_{t \neq k, %t=1}^{K}\frac{tr(\mathbf{R}_{t})tr(\mathbf{R}_{k})}{tr(\mathbf{R}_{t})} \leq  (K-1)\frac{tr(\mathbf{R}_{k}^2)}{tr(\mathbf{R}_{k})}, 
  % \implies & 1 \leq  \frac{tr(\mathbf{R}_{k}^2)}{tr(\mathbf{R}_{k})^2}, 
%\end{align}
%which is obviously a contradiction.

     %created using a simple set-up where $M=16$, $K=4$, $c_k=1 \forall k$ and each element in $\mathbf{R}_{k}$ is given by
%\begin{align}
 %   [\mathbf{R}_{k}]_{i,j}=\alpha_{\mathbf{R}_{k}}(0.1*k)^{|i-j|}
%\end{align}
%where $0.1$ is the correlation coefficient and multiplied by $k$ to obtain different coefficient of different users. Also $\alpha_{\mathbf{R}_{k}}= 10^{-4}$ is set that way as to make it comparable to the noise sweep.   

\section{IRS-User Association Optimization Problem }
\label{secoptimizedIRS}
The design of beamforming in literature is often based on two common optimization criteria--- the transmit power minimization and the maximization of the minimum (max-min) SINR. The first criterion has been the focus of several works, while the latter was dealt with in \cite{maxmin} for a single IRS system and in \cite{IRSuserassociation} for a multi IRS system with single antenna transmitter.  

To design the IRS-user association parameters for the considered multi-antenna multi-user IRSs-assisted system, we consider max-min average SINR as the performance metric to improve the performance and fairness of the system. We can formulate the max-min average SINR problem using the average SINR expression in \eqref{eqthmsimpletogetheravgSINR} as 
\begin{align}
    (P1)\hspace{0.2in}&\underset{\mathbf{\Lambda}}{\text{max}}\hspace{0.2in}\underset{k}{\text{min}}\hspace{0.2in}\bar{\gamma}_k \\
    \label{constraintlam1}
     s.t. \hspace{0.2in}&\sum_{k=1}^{K}\lambda_{l,k}=1, 1\leq l\leq L,\\\label{constraintbin}
     & \lambda_{l,k}\in \{0,1\} ,1\leq l\leq L, 1\leq k\leq K,
\end{align}
where $\boldsymbol{\Lambda} \in \mathbb{B}^{K \times L}$ denotes the binary association matrix between the $K$ users and $L$ IRSs, where each element $\lambda_{l,k}$ can take the value zero or one as mentioned in constraint \eqref{constraintbin}.%,  $\boldsymbol{\lambda}_l \in \mathbb{B}^{K \times 1}$ is a column vector which one element equalling $1$ and the rest zeros, the index of $1$ in the vector points to the associated user $k$ for the $l^{th}$ IRS, and  $\boldsymbol{\lambda}_k \in \mathbb{B}^{1 \times L}$ is a row vector of associated and non-associated $L$ IRSs for a particular user $k$. 
 The constraint \eqref{constraintlam1} limits each IRS to be associated to only one user. The binary constraint in \eqref{constraintbin} makes this problem a non-convex mixed-integer non-linear programming (MINLP) problem, which is NP-hard. We can find the optimal solution for the association parameters by exhaustive search, but the complexity is prohibitive and in the order $\mathcal{O}(K^L)$. In the next subsections, we define the search space (codebook) for the exhaustive search method and outline a low-complexity SR algorithm. \subsubsection{Exhaustive Search}
In exhaustive search, we find the IRS-user association matrix that maximizes the minimum SINR over all possible IRS-user association matrices taken from a specific codebook. To create the codebook $C \in \mathbb{B}^{K \times L \times K^L}$ that can generate all possible $K^L$   combinations of $\mathbf{\Lambda}$, we define the rules according to $(P1)$ as:\begin{enumerate}
    \item \label{item 1} Each IRS $l$ is associated to only one user $k$.
    \item $\lambda_{l,k}$ is a binary variable.
\end{enumerate}
We assign a number $N \in \mathbb{Z}^+$, where $N$ can take values from $ {1, \dots, K^L}$  to uniquely represent each matrix $\mathbf{\Lambda}_N$. We then find the base $K$ representation of $N$ and store it in a row vector $\mathbf{r}$ which is of size $(L +1) \times 1$. The first $L$ elements of $\mathbf{r}$  which can only take on values ranging from $k=0,\dots, K-1$ by definition of base $K$ conversion, map to $L$ vectors each denoted by $\boldsymbol{\lambda}_l \in \mathbb{B}^{K \times 1}$. Each element  $\mathbf{r}[l], l=1,\dots,L$  denotes the position index $i=\mathbf{r}[l]+1$ of $1$ in its corresponding $\boldsymbol{\lambda}_l$ such that $\boldsymbol{\lambda}_l[i]=1$ , where the rest of $\boldsymbol{\lambda}_l$'s elements are zeros.
%\in \mathbb{F}_K$, where $\mathbb{F}_K$ is a finite Galois field with order $K$.
%Each element $k$ in $\mathbf{r}, k=1,\dots, K$  denotes the  position index of $1$ in $\boldsymbol{\lambda}_l \in \mathbb{B}^{K \times 1} $ where the rest of its elements are all zeros, because of  \eqref{constraintlam1}. 
Thus, these $L$ $\boldsymbol{\lambda}_l$ columns are assembled as $\boldsymbol{\Lambda}_N$.

\textbf{Example:}
For clarity, we give an example with $K=2,L=2$. The number of association matrices to generate is $K^L=4$ and are shown in \eqref{eqmatrices}.
When $N=1$, the binary conversion where the most significant bit is here unconventionally the left-most bit will be $\mathbf{r}=[1, 0 ,0]$. This corresponds to having the index $(\mathbf{r}[1]+1=2)$ for one in $\boldsymbol{\lambda}_1=[0 ,1]^T$ and $(\mathbf{r}[2]+1=1)$ for one in $\boldsymbol{\lambda}_2=[1 ,0]^T$. Assembling $\boldsymbol{\lambda}_1$ and $\boldsymbol{\lambda}_2$ results in $\boldsymbol{\Lambda}_1$. We skip $N=2,N=3$ since they follow similarly. For $N=4$, the binary conversion is  $\mathbf{r}=[0, 0 ,1]$. This corresponds to the index $(\mathbf{r}[1]+1=1)$ for one in $\boldsymbol{\lambda}_1=[1 ,0]^T$ and $(\mathbf{r}[2]+1=1)$ for one in $\boldsymbol{\lambda}_2=[1 ,0]^T$. Assembling  $\boldsymbol{\lambda}_1, \boldsymbol{\lambda
}_2$ results in $\boldsymbol{\Lambda}_4$.
\begin{align}
\label{eqmatrices}
\boldsymbol{\Lambda}_1=\begin{bmatrix}
0& 1\\
    1 & 0
\end{bmatrix},
\boldsymbol{\Lambda}_2=\begin{bmatrix}
1& 0\\
    0 & 1
\end{bmatrix},
\boldsymbol{\Lambda}_3=\begin{bmatrix}
0& 0\\
    1 & 1
\end{bmatrix},
\boldsymbol{\Lambda}_4=\begin{bmatrix}
1& 1\\
    0 & 0
\end{bmatrix}
\end{align}

%need to transform the problem into a simpler and already solved  base conversion problem. First, 

     \subsubsection{SR Algorithm}
      \begin{algorithm}
\label{algorithm1}
\SetAlgoLined
\KwResult{Optimized IRS-user association matrix $\mathbf{\Lambda}^*$;}
 Initialize $\mathbf{\Lambda}$ based on nearest distance rule\;
 Initialize $\mathbf{v}^{k_l}_l$ using \eqref{eqirsphase}; Compute  $\bar{\gamma}_k$s in \eqref{eqthmsimpletogetheravgSINR} based on $\mathbf{\Lambda}$ and $\mathbf{v}^{k_l}_l$\;
Set the iteration number $i=1$ and set $state =true$ \;
 \While{state is true}{
  $k$ =  $\text{arg min}_k(\bar{\gamma}_k)$  (bottleneck user), $\bar{\gamma}_{\text{min}}(i)=\bar{\gamma}_{k}$\;
   \For{$l=1:L$}{
  \If{$\lambda_{l,k}=0$}{
  $\lambda_{l,k} =1$, $ \lambda_{l,j} =0$,  $j \neq k$, where $j$ is the index of previously associated user of IRS $l$\;
   Update $\bar{\gamma}_k$s, $k=1,\dots, K$\;
	Find $\tilde{\gamma}(l)=\text{min}_k (\bar{\gamma}_k)$\;
  Reset $\lambda_{l,k} =0$, $ \lambda_{l,j} =1$\;
   }
    }% \EndFor
     Find $\bar{l}$ =  $\text{arg max}_l(\tilde{\gamma}(l))$, i.e. the best IRS to improve the minimum SINR\;
      $\lambda_{\bar{l},k} =1$, $ \lambda_{\bar{l},{j}} =0$, where $j$ is the index of previously associated user of IRS $\bar{l}$\;       
       Update $\bar{\gamma}_k$, $k=1,\dots, K$ and find $\bar{\gamma}_{\text{min}}(i+1)=\text{min}_k (\bar{\gamma}_k)$\;
         \If{$\bar{\gamma}_{\text{min}}(i+1)<\bar{\gamma}_{\text{min}}(i)$ }{  $state=false$,     $\lambda_{\bar{l},k} =0$, $ \lambda_{\bar{l},{j}} =1$\; }   $i=i+1$\;}
        \caption{SR Algorithm to Solve $(P1)$}
 \label{algorithm1}
\end{algorithm}
     To observe the gains yielded by optimizing IRS-user associations with low computational complexity, we outline the SR algorithm to maximize the minimum SINR, which was first proposed in \cite{IRSuserassociation}. First, we initialize the association matrix $\boldsymbol{\Lambda}$ using an appropriate criteria, like associating IRSs based on the minimum distance to the users. Under this initialization, we compute the average SINRs for all users and find the weakest (or the bottleneck user) user, which has the lowest average SINR $\bar{\gamma}_k$. After that, we search for an IRS that will increase the average SINR of this bottleneck user while not decreasing the overall system's minimum SINR. We iterate in this manner until there can be no further improvement in the minimum average SINR of the system. The SR algorithm is summarized in Algorithm \ref{algorithm1}. This method has a complexity in the order of $ \mathcal{O}(L)$. % which is better than the complexity of the well-known branch and bound algorithm for mixed integer linear programming problems. In fact, branch and bound algorithm can reach a complexity as high as that of exhaustive search, i.e. $\mathcal{O}(K^L)$. 
     The minimum average SINR performance under SR algorithm matches closely to that under the optimal solution yielded by exhaustive search as shown in the simulation results.

\section{ Numerical Simulations and Discussion}
\label{secresults}
\begin{figure}[H]
    \centering
    \subfloat[Simulation parameters.]{
\scalebox{0.6}{\begin{tabular}{|l|l|}
\hline
  \textbf{Parameter} & \textbf{Value} \\ 
\hline
\textbf{Array parameters:} & \\
\hline
%Carrier frequency & $2.5$ GHz \\
 BS configuration & Uniform linear array\\
IRS configuration & Uniform planar array\\
 Antenna gain & 5dBi\\
$d_{BS}$, $d_{IRS}$& $0.5\lambda$\\
Noise level & $-60$\rm{dBm} \\
 \hline
\textbf{Path Loss:} & \\
\hline
Model & $\frac{10^{-C/10}}{d^{\alpha}}$ \\
$C$ (Fixed loss at $d=1$m)  & $25$\rm{dB} ($\beta_{1}$), $30$\rm{dB} ($\beta_{2,k},\beta_{d,k}$)\\
$\alpha$ (Path loss exponent) & $2.2$ ($\beta_{1}$), $3.67$ ($\beta_{2,k},\beta_{d,k}$)\\
\hline
\textbf{Penetration Loss:} & \\
\hline
$(\mathbf{h}_{d,k},\mathbf{h}_{2,l,k})$ & ($20$\rm{dB}, $5$\rm{dB})\\ 
\hline
\textbf{System Dimensions:} & \\
\hline
$(L,K,M)$ & $(8,4,16)$ \\

\hline
%\textbf{Deployment Scenario} & \\
%\hline
%IRSs & arc of radius $100m$ \\
%Users & arc of radius $85m$ \\
%User $2$ & at radius $130m$ \\
%\hline
\end{tabular}}
\label{Tabsims}
}
\label{tabfigu}
  \subfloat[No Association Set-up]{
	\begin{minipage}[c][1\width]{
	   0.51\textwidth}
	   \centering
	   \includegraphics[width=0.9\textwidth]{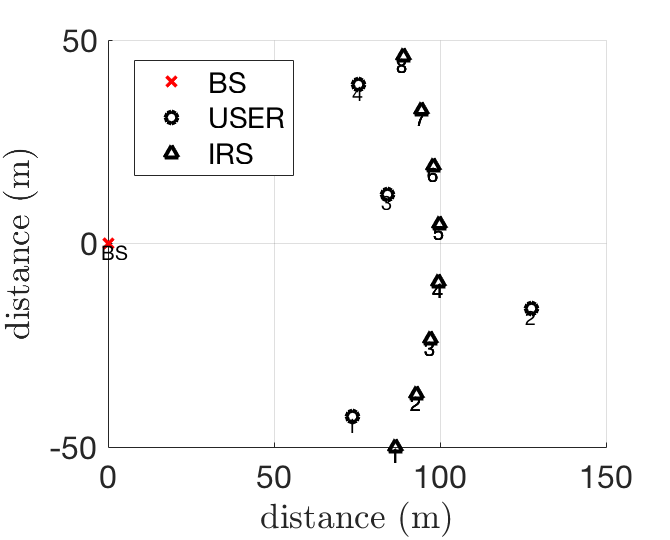}%{images/lastmin_images/systemirstriangle.png}
	   \label{fig:noassoc}
	\end{minipage}}
 \hfill 
\caption{Deployment scenario and parameters.}
\end{figure}

\begin{figure}[!t]
\centering
\begin{minipage}[b]{0.44\linewidth}
\centering %width=2.85in, height=1.9in]
\includegraphics[scale=0.24]{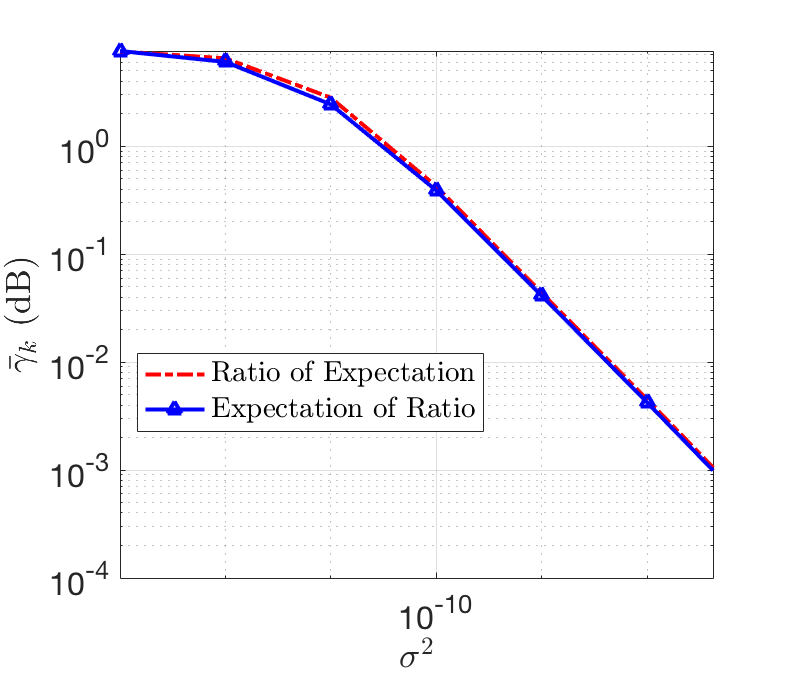}
\caption{Tight approximation of $\bar{\gamma}_k$}
    \label{figtight}
\end{minipage}
\hspace{0.5cm}
\begin{minipage}[b]{0.45\linewidth}
\centering
\includegraphics[scale=0.24]{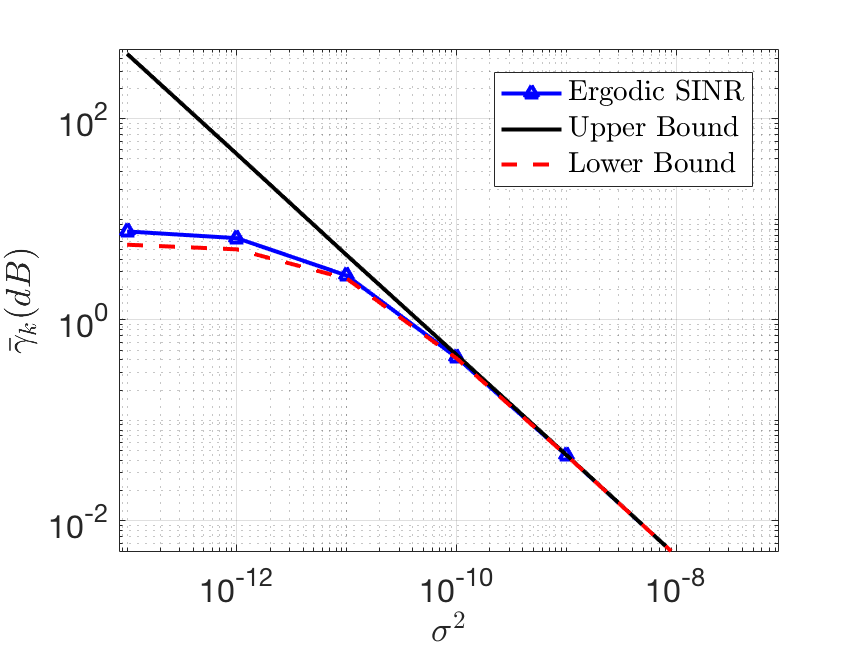}%{images/lastmin_images/ergodicwithupper.png}
\caption{Upper and lower bounds of $\bar{\gamma}_k$.}
\label{figupperlower}
\end{minipage}
\end{figure}

The parameter values for the simulation results in this section are tabulated in Table \ref{Tabsims}. We consider the following deployment. Denoting by $(x,y)$ the Cartesian coordinates, the BS is located at $(0,0)m$, the $L$ IRSs are deployed on an arc of radius $100m$ with respect to the BS, and the $K$ users are distributed on an arc of radius $85m$ with one user  (numbered 2) is set further away at a radius of $130m$. This distributed IRSs deployment is illustrated in Fig. \ref{fig:noassoc}  and the user $2$ is deliberately positioned further away to highlight how the proposed IRS-user association algorithm helps the edge (or bottleneck) users.
The path loss factors are computed at 2.5 GHz carrier frequency for the 3GPP Urban Micro (UMi) scenario from TR36.814 (also found in Section V \cite{nadeem2020intelligent}). The LoS channel model was used for $\mathbf{H}_{1,l}$ and the non-LOS (NLOS) channel model was used to generate path loss factors for $\mathbf{h}_{2,l,k}$ and $\mathbf{h}_{d,k}$, where $d$ in the path loss expression $\frac{10^{-C/10}}{d^{\alpha}}$ denotes the Euclidean distance between different nodes. Higher penetration loss is considered for the direct link due to obstacles in the environment, which can be avoided by deploying IRSs. The first figure,  Fig. \ref{figtight}, validates the expression of $\bar{\gamma}_k$ in Lemma \ref{lemavg}  by plotting it against the noise variance $\sigma^2$.  It depicts that the expectation of the ratio in \eqref{eqinstSINR2} to formulate the average SINR expression  can be very well approximated by the ratio of expectation as done in \eqref{eqsinravg1}.

 Fig. \ref{figupperlower} plots the average SINR expression given in Theorem \ref{theoremsinravg} as well as its lower bound and upper bound  given in Corollary \ref{corortho} as a function of noise variance $\sigma^2$. We see  the average SINR to be well bounded by the two bounds with the upper bound becoming very tight as $\sigma^2$ increases since the system becomes noise limited. The upper bound is linear in $\sigma^2$ since it sets the interference to zero by assuming the correlation matrices of the users to be orthogonal.

\begin{figure}[H]
\centering
  \subfloat[Nearest Rule]{
%	\begin{minipage}[c][1\width]{
	  % 0.3\textwidth}
	   \centering
	   \includegraphics[width=0.3\textwidth]{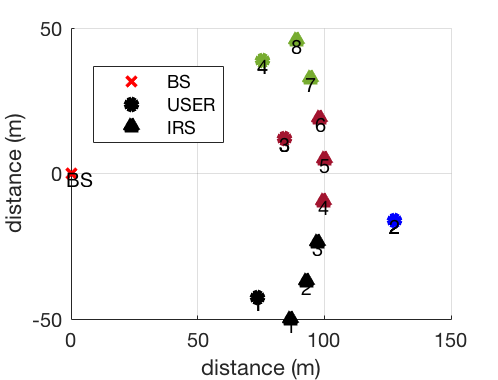}%{images/images/images_Ch45/cent_dist_NR.jpg}
	   \label{fig:nrassoc}
%	\end{minipage}
}
  \subfloat[Successive Refinement ]{
%	\begin{minipage}[c][1\width]{
%	   0.3\textwidth}
	   \centering
	   \includegraphics[width=0.3\textwidth]{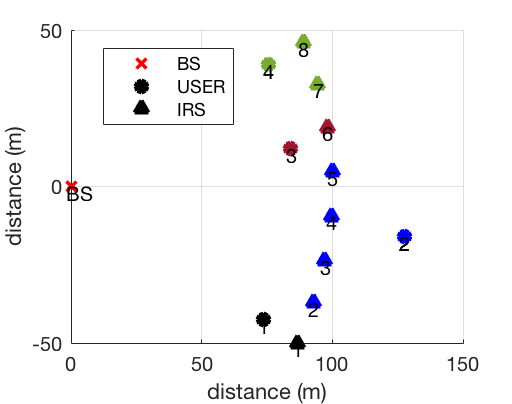}%%{images/images/images_Ch45/cent_dist_SR.jpg}
	   \label{fig:srassoc}
%	\end{minipage}
}\subfloat[Exhaustive Search ]{
%	\begin{minipage}[c][1\width]{
%	   0.3\textwidth}
	   \centering
	   \includegraphics[width=0.32\textwidth]{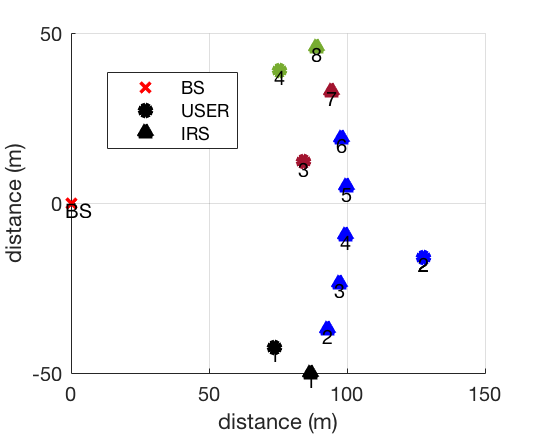}	   \label{fig:exhaustasso}
%	\end{minipage}
}
\caption{Association shown by color scheme. An IRS and a user are shown to be associated by having the same color.}\label{fig:three graphs}
\end{figure}

\begin{figure}[!t]
\centering
\begin{minipage}[b]{0.43\linewidth}
\centering %width=2.85in, height=1.9in]
\includegraphics[scale=0.2]{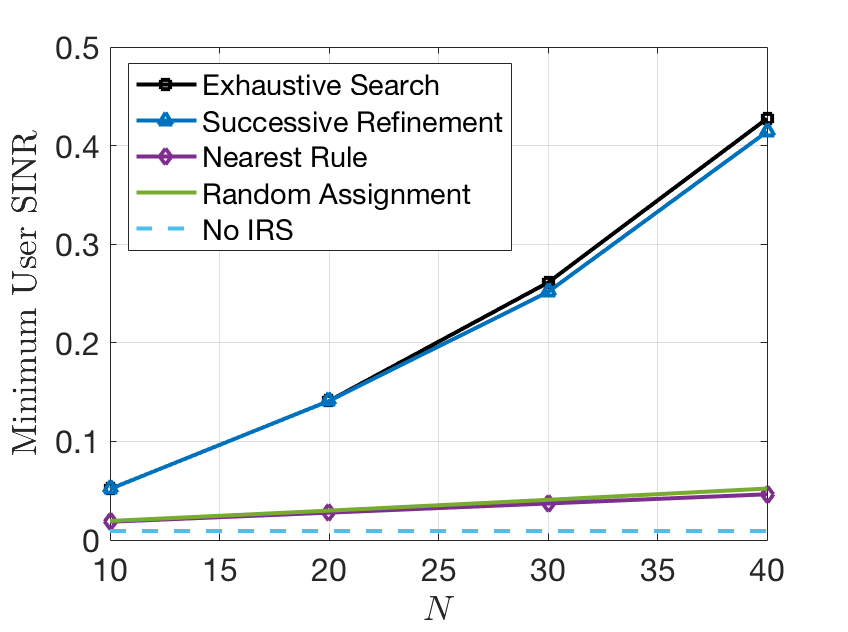}
    \caption[Minimum user SINR against $N$]{Minimum user SINR vs $N$.}
      \label{fig:EXHAUSTn}
\end{minipage}
\hspace{0.5cm}
\begin{minipage}[b]{0.4\linewidth}%  \includegraphics[scale=0.28]{L_monte_Carlo_1.png}
\centering \includegraphics[scale=0.17]{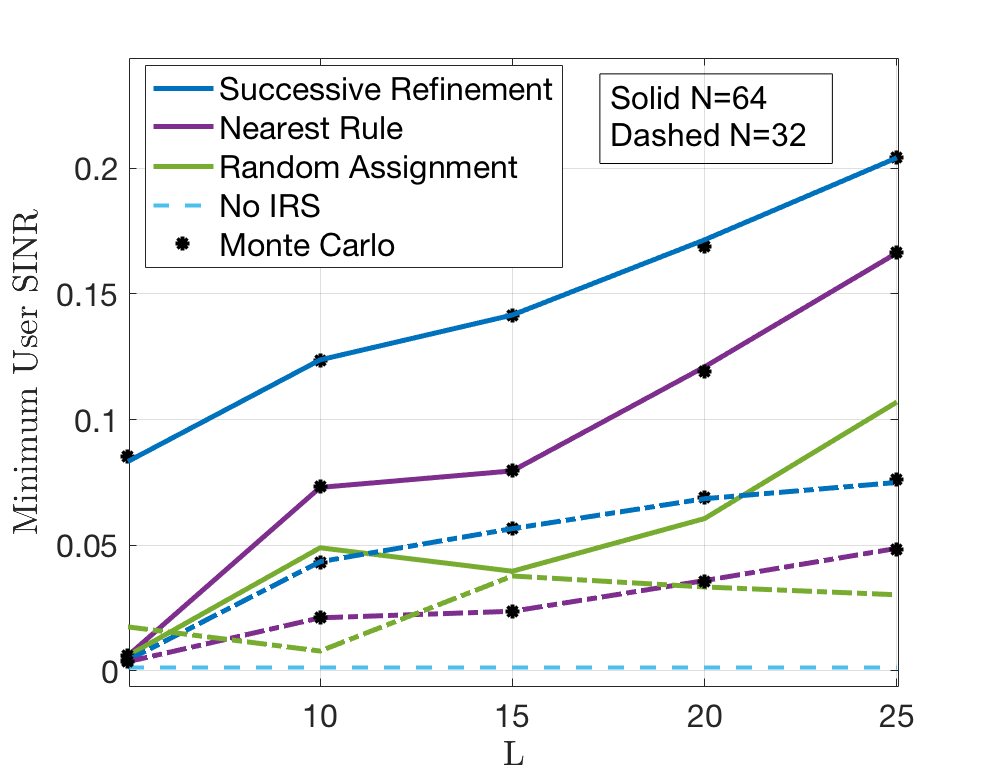}
    \caption[Minimum user SINR against $L$]{Minimum user SINR vs $L$. }
    \label{fig:againstL}
\end{minipage}
\end{figure}

Next we study how the users in Fig. \ref{fig:noassoc} are associated with different IRSs under different IRS-user association methods. In Fig. \ref{fig:nrassoc}, the associations are updated based on nearest rule, where each IRS is assigned to the user that has the shortest distance to it. On the other hand,  Fig. \ref{fig:srassoc} shows the IRS-user association resulting from the outlined SR Algorithm \ref{algorithm1}, which focuses on maximizing the SINR of the bottleneck user. It is clear from the deployment in Fig. \ref{fig:noassoc} that user $2$ is the bottleneck user and we can see in Fig. \ref{fig:srassoc} that SR algorithm managed to assign more IRSs to this user. We will see later in Fig. \ref{fig:EXHAUSTn} that by doing so, the system achieves a higher minimum average SINR. Fig. \ref{fig:exhaustasso} shows the result under exhaustive search IRS-user association. Again more IRSs are assigned to the second user is determined in the deployment to be the bottleneck user.
Plotting the minimum user average SINR with average SINR derived in \eqref{eqthmsimpletogetheravgSINR} against $N$ in Fig. \ref{fig:EXHAUSTn}, we notice that the SR  algorithm explained in Algorithm \ref{algorithm1} achieves a close performance to that of exhaustive search. Exhaustive search has complexity in the order of $\mathcal{O}(K^L)$, and would therefore not scale well with the number of users and IRSs in the system. Hence the SR algorithm, which scales linearly with $L$, is an effective method to (nearly optimally) solve the IRS-user association problem. We also observe from Fig. \ref{fig:EXHAUSTn} that the minimum user SINR improves with increasing  the number of elements $N$ but not in the order of $N^2$. Generally for single user systems where there is no interference, deploying IRSs  achieves gains in the order $N^2$ in the receiver's SNR  which is a combination of the array and passive beamforming gains \cite{maxmin}, \cite{mythsquestions}. The reason we do not see an $N^2$ gain in multi-user systems is when we increase $N$ we not only increase the channel gain at the user but also the interference experienced.

Also note that the nearest distance rule and random assignment yield a much lower minimum average SINR, which is attributed to the fact that they do not necessarily help the bottleneck user. However, a distributed IRSs-assisted system with nearest distance rule and random IRS-user assignment does perform better than a system with no IRS because as we mentioned before, even the non-associated IRSs will contribute  to the bottleneck user's channel gain which is seen in Lemma \ref{lemmanormhk2} and the discussion underneath it.

Next we study the minimum average SINR against $L$ in Fig. \ref{fig:againstL}. We plot the Monte-Carlo simulated average SINR in \eqref{eqsinravg1} as well as the derived expression  \eqref{eqthmsimpletogetheravgSINR} for the average SINR in Theorem \ref{theoremsinravg}. The Monte-Carlo simulated values are averaged over 1000 channel realizations and are shown to match the theoretical expression accurately thereby validating Theorem \ref{theoremsinravg}. Since exhaustive search does not scale well with increasing $L$, we did not include it in this plot. However, it has already been established in the previous figure that the solution yielded by SR performs very close to that yielded by exhaustive search which is optimal. Increasing the number of IRSs is beneficial as expected, since there are more IRSs to be assigned to the users resulting in higher, passive beamforming gains. %Performance can be further improved if the locations of the IRSs are carefully selected as to broaden coverage. However, that is not the focus of this work. 
The solid lines in the figure represent a doubled number of IRS elements than that represented by the dashed lines. We can see that doubling $N$ improves the minimum user SINR as expected. Moreover, we see that nearest distance rule assignment increases the minimum user SINR but at a slower rate than SR. Meanwhile, random assignment may increase the minimum user SINR or not as depicted, since the IRSs can be associated randomly in a detrimental manner to the minimum user SINR.
\begin{figure}[H]
    \centering
    \includegraphics[scale=0.2]{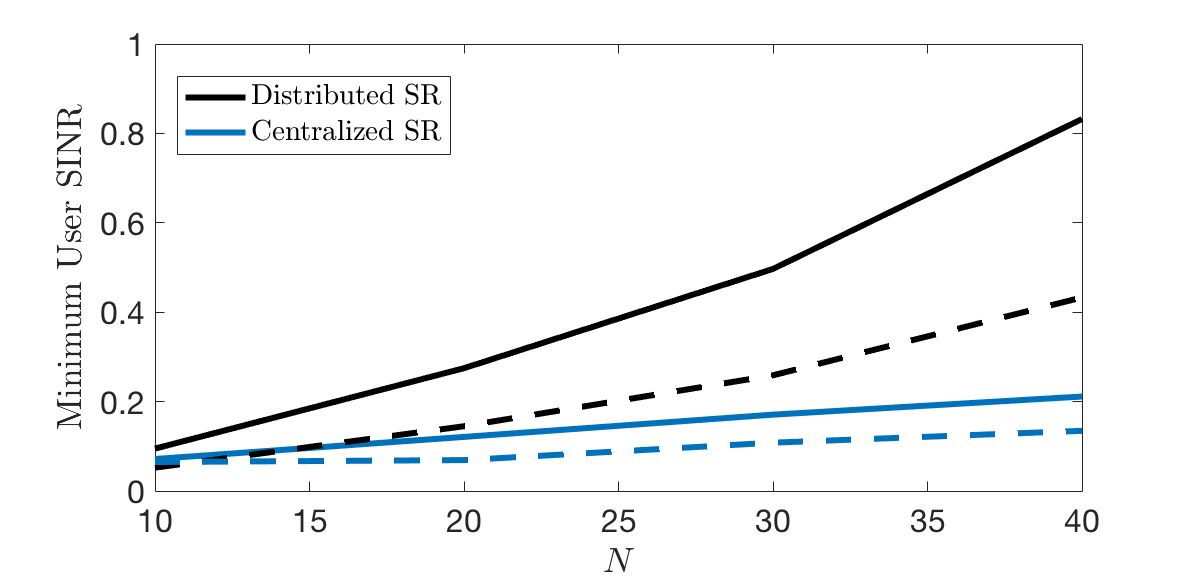}
    \caption[Minimum user SINR for centralized and distributed IRSs scenarios against N]{Minimum user SINR against $N$ under centralized and distributed IRSs deployment scenarios. Solid lines represent $L=16$, dashed lines represent $L=8$.}
    \label{fig:centralized}
\end{figure}
Fig. \ref{fig:centralized} depicts the minimum user SINR  when the IRSs are distributed as opposed to having the IRSs as one central unit. Centralized deployment of IRSs compacts the $L$ IRSs, which can be effectively represented by a single large IRS  with $NL$ elements. Distributed IRSs assisted system performs better than the centralized IRS assisted system because of the spatial diversity that the distributed placement offers.  Especially under LoS BS-IRS propagation, the BS-IRS channel is almost certain to be rank-deficient since the IRS is in the far-field of the BS, which lessens the degrees of freedom offered by the channel. Having the IRSs distributed increases the degrees of freedom to at least $L$, resulting in better minimum SINR performance.   
 \begin{figure}[H]
\centering
  \subfloat[Distributed IRSs.]{
%	\begin{minipage}[c][1\width]{
	  % 0.3\textwidth}
	   \centering
	   \includegraphics[width=0.44\textwidth]{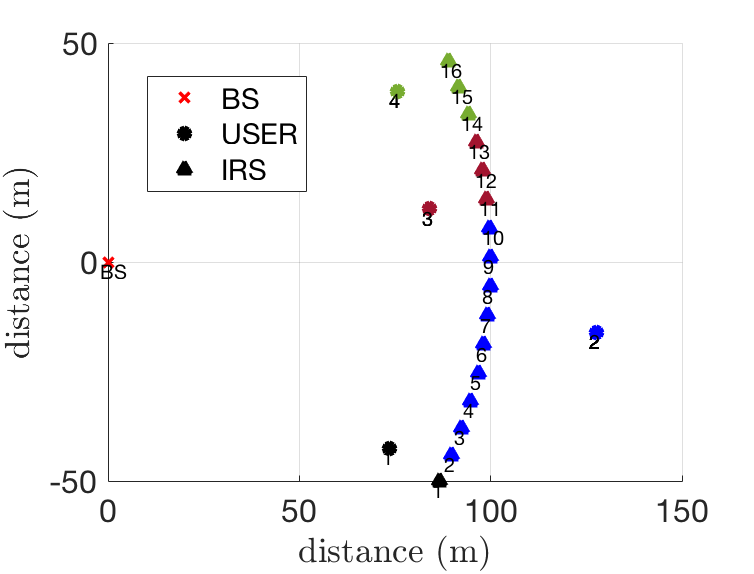}%{images/png/sinr_dis_15l.png}
    \label{fig:dist}
%	\end{minipage}
}
	\hfill
  \subfloat[Centralized IRSs]{
%	\begin{minipage}[c][1\width]{
%	   0.3\textwidth}
	   \centering
 \includegraphics[scale=0.28]{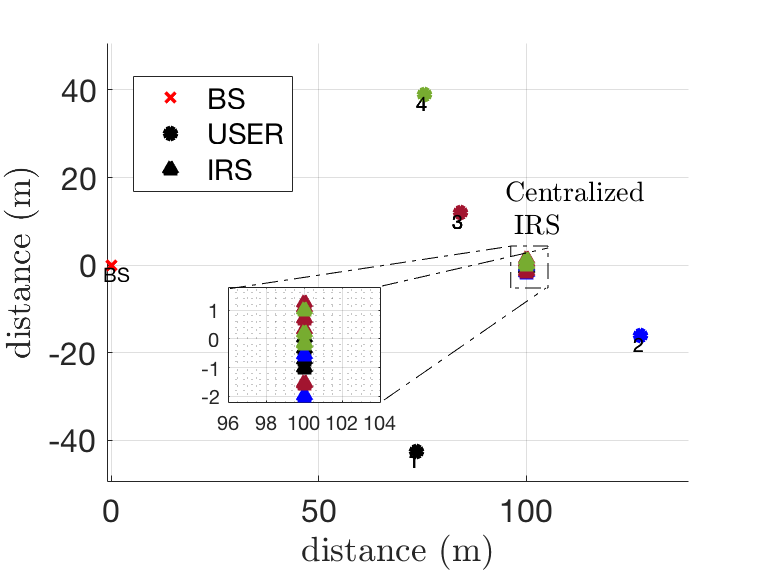}%{images/png/sinr_Centralized_15L.png}
    \label{fig:cent}}
        \caption{IRSs-users association patterns under distributed and centralized deployments. }\label{fig:three graphs}
\end{figure}
%To visualize this deployment scenario for $L=16$, we plot in Fig. \ref{fig:dist} and Fig. \ref{fig:cent} the SR optimized association for both the distributed and centralized IRSs, respectively. One can see that the centralized IRSs are not majorly assigned to the bottleneck user which was assumed to be $2$, since there are high penetration losses between the BS and the other users as well as through the direct link $\mathbf{h}_{d,k}$s. Hence, the lack of nearby IRSs is severe on the system performance, and the minimum user SINR bounces back and forth between the users furthest from the IRSs in each iteration of SR algorithm largely due to this deployment scenario.
%%%%%See if this is better%%%%%%%%%%%%%%%%%%%%%%%%%%%%%%%%%%%
Next we study the way the users are associated with the IRSs when the IRSs are distributed versus when all IRSs are located together. Considering $L=16$, we plot in Fig. \ref{fig:dist} and Fig. \ref{fig:cent} the SR optimized association for the distributed and centralized IRSs scenarios respectively. One can see that when IRSs are located together, almost all the users have a larger distance to the centralized IRSs and will experience higher path loss resulting in the SR algorithm to balance the IRSs between all users. Whereas in the distributed case, the IRSs are better distributed to cover all users and the SR algorithm will assign most IRSs to the bottleneck user (user $2$). The lack of nearby IRSs to all users has an adverse impact on the system performance, and the minimum user SINR deteriorates under the centralized deployment.

\section{Conclusion}
\label{chconc}

\label{conclusion}
The work considered a distributed IRSs assisted multi-user MISO system, where the IRSs are associated with users in an optimized manner. We derived a tractable average SINR expression using statistical tools under MRT precoding, optimized IRS reflect beamforming, and arbitrary IRS-user association parameters. The expression was used to formulate and solve a max-min average SINR problem to optimize the IRS-user association parameters using a low-complexity SR algorithm. Simulation results validated the average SINR expression and studied the effect of increasing the number of IRSs, the number of elements in each IRS, as well as changing the physical distribution of the IRSs and IRS-user association pattern. In particular, the results show that  that the minimum user SINR gets quadrupled when the IRSs are deployed in a distributed manner as opposed to centralized manner. 
The sub-optimal SR algorithm is shown to perform closely to the optimal solution given by exhaustive search. 
An important future direction will be to extend the theoretical analysis for average SINR to imperfect CSI scenario as well as multi-cell systems. 
%One other possible direction to take is to derive the average SINR without first imposing a solution for reflect beamforming vector at each IRS, and then optimize over both association parameters and reflect beamforming.
%\appendix 
\begin{appendices}
\section{Proof of Lemma \ref{lemmanormhk2}}
\label{appendixIEEEproofchannelgainsq}
We wish to show the expectation of the squared norm of the channel given as
        \begin{align}
&\mathbb{E}[\|{\mathbf{h}}_{k}(\boldsymbol{\lambda_{k}})\|^2]= E_1+E_2+E_3+E_4+E_5+E_6,\\
    &\text{where, } \hspace{.1in} E_1=\mathbb{E}[\mathbf{h}_{d,k}^H\mathbf{h}_{d,k}]=M\beta_{d,k},\\&E_2=\mathbb{E}[2\sum_{l=1}^{L}\lambda_{l,k}\mathbf{v}_{l}^{{k_l}^H}  \mathbf{H}_{0,l,k}^H\mathbf{h}_{d,k}],\\&E_3=\mathbb{E}[\sum_{l=1}^{L}\sum_{\bar{l}=1}^{L}\lambda_{l,k}\lambda_{\bar{l},k}\mathbf{v}_{l}^{{k_l}^H}  \mathbf{H}_{0,l,k}^H\mathbf{H}_{0,\bar{l},k}\mathbf{v}_{\bar{l}}^{{k_{\bar{l}}}}], \\ \label{eqE4}&E_4=\mathbb{E}[\sum_{l=1}^{L}\sum_{\bar{l}=1}^{L}(1-\lambda_{l,k})(1-\lambda_{\bar{l},k})\mathbf{v}_{l}^{k_l^H}  \mathbf{H}_{0,l,k}^H\mathbf{H}_{0,\bar{l},k}\mathbf{v}_{\bar{l}}^{{k_{\bar{l}}}}],\\&E_5=\mathbb{E}[2\sum_{l=1}^{L}(1-\lambda_{l,k})\mathbf{v}_{l}^{k_l^H}  \mathbf{H}_{0,l,k}^H\mathbf{h}_{d,k}],\\&E_6=\mathbb{E}[2\sum_{l=1}^{L}\sum_{\bar{l}=1}^{L}(1-\lambda_{\bar{l},k})\lambda_{{l},k}\mathbf{v}_{l}^{{k_l}^H}  \mathbf{H}_{0,l,k}^H\mathbf{H}_{0,\bar{l},k}\mathbf{v}_{\bar{l}}^{{k_{\bar{l}}}}],
    \end{align}     
where $\mathbf{v}_{l}^{k_l}$ defined in \eqref{eqirsphase} and  ${\mathbf{h}}_{k}(\boldsymbol{\lambda}_{k})$ defined in \eqref{eqvlambda}.  First we work on $E_2$, which is the inner product of the cascaded IRS channel and the direct channel with the optimized IRS configuration $\mathbf{v}_{l}^{k_l}$ where $k_l=k$. We obtain
  \begin{align}
    &E_2=2\sum_{l=1}^{L}\lambda_{l,k}\mathbb{E}[\mathbf{v}_{l}^{{k_l}^H}  \mathbf{H}_{0,l,k}^H\mathbf{h}_{d,k}] = 2\sum_{l=1}^{L}\lambda_{l,k}\mathbb{E}[\sum_{n=1}^{N}e^{j\angle{{h}_{2,l,k,n}{b}_{l,n}^*}}e^{j\angle{\mathbf{h}_{d,k}^H\mathbf{a}_l}}{h}_{2,l,k,n}^*{b}_{l,n}\mathbf{a}_l^H\mathbf{h}_{d,k}]\\&= 2\sum_{l=1}^{L}\lambda_{l,k}\mathbb{E}[\sum_{n=1}^{N}e^{j(\angle{{h}_{2,l,k,n}{b}_{l,n}^*}-\angle{{h}_{2,l,k,n}{b}_{l,n}^*})}e^{j(\angle{\mathbf{h}_{d,k}^H\mathbf{a}_l}-\angle{\mathbf{h}_{d,k}^H\mathbf{a}_l})}|{h}_{2,l,k,n}^*{b}_{l,n}||\mathbf{a}_l^H\mathbf{h}_{d,k}|]\\ \label{eqrayleighe2}&=2\sqrt{\beta_{1,l}}\sum_{l=1}^{L}\lambda_{l,k}\mathbb{E}[\sum_{n=1}^{N}| {h}_{2,l,k,n}|\cdot|\mathbf{a}_{l}^H\mathbf{h}_{d,k}|]= 2\sqrt{\beta_{1,l}\beta_{d,k}\beta_{2,l,k}}\sum_{l=1}^{L}\lambda_{l,k}\sum_{n=1}^{N}\frac{\pi}{4}\sqrt{\mathbf{a}_{l}^H\mathbf{a}_{l}}\\\label{eqE2final}&=2\sqrt{\beta_{1,l}\beta_{d,k}\beta_{2,l,k}}\sum_{l=1}^{L}\lambda_{l,k}\frac{\pi}{4}\sqrt{M}N,%=\\&2\sum_{l=1}^{L}\lambda_{l,k}\frac{\pi}{4}\sqrt{tr\mathbf{I}_{N})}\sqrt{tr(\mathbf{H}_{1,l}^H\mathbf{H}_{1,l})}
        \end{align}
   where $\mathbf{H}_{0,l,k}$ is given  in  \eqref{eqholk}, ${h}_{2,l,k}(n)$ is the $n^{th}$ element in $\mathbf{h}_{2,l,k}$, and \eqref{eqrayleighe2} follows since $|{h}_{2,l,k}(n)|$ and $|\mathbf{a}_{l}^H\mathbf{h}_{d,k}|$ are statistically independent Rayleigh distributed random variables whose mean values are given as $\sqrt{\pi\beta_{2,l,k}}/2$ and  $\sqrt{\pi\beta_{dk}\mathbf{a}_{l}^H\mathbf{a}_{l}}/2$, respectively. 
	
   Next we work on $E_3$  where $k_l=k$ as follows
      \begin{align}
      \label{eqe3all}
    &E_3=\mathbb{E}[\sum_{l=1}^{L}\sum_{\bar{l}=1}^{L}\lambda_{l,k}\lambda_{\bar{l},k}\mathbf{v}_{l}^{{k_l}^H}  \mathbf{H}_{0,l,k}^H\mathbf{H}_{0,\bar{l},k}\mathbf{v}_{\bar{l}}^{{k_{\bar{l}}}}] \\&=\sum_{l=1}^{L}\lambda_{l,k}\mathbb{E}[\mathbf{v}_{l}^{{k_l}^H}  \mathbf{H}_{0,l,k}^H\mathbf{H}_{0,{l},k}\mathbf{v}_{l}^{k_l}] +\sum_{l=1}^{L}\sum_{\bar{l}\neq l}^{L}\lambda_{l,k}\lambda_{\bar{l},k}\mathbb{E}[\mathbf{v}_{l}^{{k_l}^H}  \mathbf{H}_{0,l,k}^H\mathbf{H}_{0,\bar{l},k}\mathbf{v}_{\bar{l}}^{{k_{\bar{l}}}}] %,\\&=\sum_{l=1}^{L}\lambda_{l,k}\mathbb{E}[\mathbf{v}_{l}^{{k_l}^H}  \mathbf{H}_{0,l,k}^H\mathbf{H}_{0,{l},k}\mathbf{v}_{l}^{k_l}]=\sum_{l=1}^{L}\lambda_{l,k}\mathbb{E}[\mathbf{e}^{j\angle\mathbf{h}_{d,k}^H\mathbf{H}_{0,l,k}}  \mathbf{H}_{0,l,k}^H\mathbf{H}_{0,{l},k}\mathbf{e}^{j\angle\mathbf{H}_{0,l,k}^H\mathbf{h}_{d,k}}], 
    \\&=\sum_{l=1}^{L}\lambda_{l,k}\mathbb{E}[\|\mathbf{H}_{0,{l},k}\mathbf{v}_{l}^{k_l}\|^2],
        \end{align}
      where the term with the  sum  $l \neq \bar{l}$ is  zero due to independence between the different $\mathbf{h}_{2,l,k}$, $l=1,\dots, L$.    Now, we can use a result for CGQF \cite{cgqf} to obtain 
        \begin{align}
         E_3&=\sum_{l=1}^{L}\lambda_{l,k}\mathbb{E}[\mathbf{e}^{j\angle\mathbf{h}_{d,k}^H\mathbf{H}_{0,l,k}}  \mathbf{H}_{0,l,k}^H\mathbf{H}_{0,{l},k}\mathbf{e}^{j\angle\mathbf{H}_{0,l,k}^H\mathbf{h}_{d,k}}]=\sum_{l=1}^{L}\lambda_{l,k}\mathbb{E}[\tilde{\mathbf{v}}_{l}^{{k_l}^H} \mathbf{A}\tilde{\mathbf{v}}_{l}^{{k_l}}],
        \end{align}
        where $\tilde{\mathbf{v}}_{l}^{{k_l}}=\text{diag}(|\mathbf{h}_{2,l,k}|)\mathbf{e}^{j\angle\mathbf{H}_{1,l}^H\mathbf{h}_{d,k}}$, and
        \begin{align}
        \label{eqA}
          \mathbf{A}=\mathbf{H}_{1,l}^H\mathbf{H}_{1,{l}}=\beta_{1,l}\|\mathbf{a}_l\|^2\mathbf{b}_l\mathbf{b}_l^H
        \end{align} is a symmetric, deterministic matrix.
        We can now write 
    \begin{align}
    \label{eqquadradticformappendix}
       E_3 &=\sum_{l=1}^{L}\lambda_{l,k}\mathbb{E}[\tilde{\mathbf{v}}_{l}^{{k_l}^H} \mathbf{A}\tilde{\mathbf{v}}_{l}^{{k_l}}]=\sum_{l=1}^{L}\lambda_{l,k}\big(tr(\mathbf{A}\boldsymbol{\Sigma}_{\tilde{\mathbf{v}}_{l}^{{k_l}}})+\boldsymbol{\mu}_{\tilde{\mathbf{v}}_{l}^{{k_l}}}^H\mathbf{A}\boldsymbol{\mu}_{\tilde{\mathbf{v}}_{l}^{{k_l}}}\big),
        \end{align}   
        where $\boldsymbol{\Sigma}_{\tilde{\mathbf{v}}_{l}^{{k_l}}}=\mathbb{E}[\tilde{\mathbf{v}}_{l}^{{k_l}} \tilde{\mathbf{v}}_{l}^{{k_l}^H}]$ and  $\mu_{\tilde{v}_l^{k}}=  \mathbb{E}[\tilde{\mathbf{v}}_{l}^{{k_l}}]$. Next, we express $\boldsymbol{\Sigma}_{\tilde{\mathbf{v}}_{l}^{{k_l}}}$ using Lemma \ref{lemdad}.  
       
         \begin{lemma}
         \label{lemdad}
            The expectation of $\mathbf{D}\mathbf{A}\mathbf{D}$ where $\mathbf{D} \in \mathbb{C}^{S \times S}$ is a random diagonal matrix  and $\mathbf{A} \in \mathbb{C}^{S \times S}$ is a symmetric, deterministic matrix is given by
              \begin{align}
     &\mathbb{E}[[\mathbf{D}\mathbf{A}\mathbf{D}]_{i,j}]=\mathbb{E}[\mathbf{d}_i^T\mathbf{A}_{i,j}\mathbf{d}_j], \forall i, j
            \end{align}
            where $\mathbf{d}_{i}$ is the $i^{th}$ column in $\mathbf{D}$ and
              \begin{align}
      &\mathbb{E}[\mathbf{d}_i^T\mathbf{A}_{i,i}\mathbf{d}_i]=\mathbb{E}[\mathbf{d}_i^T\mathbf{d}_i]\mathbf{A}_{i,i}=\sum_{s=1}^S\mathbb{E}[{d}_{i,s}^2]\mathbf{A}_{i,i}=\sum_{s=1}^S(\text{Var}({d}_{i,s})+\mathbb{E}[{d}_{i,s}]^2)\mathbf{A}_{i,i}, \\
      &\mathbb{E}[\mathbf{d}_i^T\mathbf{A}_{i,j}\mathbf{d}_j]=\mathbb{E}[\mathbf{d}_i^T]\mathbf{A}_{i,j}\mathbb{E}[\mathbf{d}_j], \hspace{.1in} i \neq j.
            \end{align}
                     
         \end{lemma}
             
  Using Lemma \ref{lemdad}, the covariance matrix  $\boldsymbol{\Sigma}_{\tilde{\mathbf{v}}_{l}^{{k_l}}}$ is found to be
         \begin{align}\label{eqsigmavkfirst}
       &\boldsymbol{\Sigma}_{\tilde{\mathbf{v}}_{l}^{{k_l}}}=
        \mathbb{E}[\tilde{\mathbf{v}}_{l}^{{k_l}} \tilde{\mathbf{v}}_{l}^{{k_l}^H}]=    \mathbb{E}[\text{diag}(|\mathbf{h}_{2,l,k}|) \mathbf{e}^{j\angle\mathbf{H}_{1,l}^H\mathbf{h}_{d,k}}\mathbf{e}^{j\angle\mathbf{h}_{d,k}^H\mathbf{H}_{1,l}}\text{diag}(|\mathbf{h}_{2,l,k}|)]\\ 
    &=\mathbb{E}_{\mathbf{h}_{2,l,k}}[\text{diag}(|\mathbf{h}_{2,l,k}|)\mathbb{E}_{\mathbf{h}_{d,k}}[ \mathbf{e}^{j\angle\mathbf{b}_{l}\mathbf{a}_{l}^H\mathbf{h}_{d,k}}\mathbf{e}^{j\angle\mathbf{h}_{d,k}^H\mathbf{a}_{l}\mathbf{b}_{l}^H}]\text{diag}(|\mathbf{h}_{2,l,k}|)]\label{eqsigmalemma}
        %\\&=\mathbb{E}_{\mathbf{h}_{2,l,k}}[\text{diag}(|\mathbf{h}_{2,l,k}|) \mathbf{e}^{j\angle\mathbf{b}_{l}}\mathbf{e}^{j\angle\mathbf{b}_{l}^H}]\text{diag}(|\mathbf{h}_{2,l,k}|)],\\&=\mathbb{E}_{\mathbf{h}_{2,l,k}}[\text{diag}(|\mathbf{h}_{2,l,k}|) \mathbf{e}^{j\angle\mathbf{H}_{1,l}^H\mathbf{H}_{1,l}}\text{diag}(|\mathbf{h}_{2,l,k}|)]
        \\&=
        (1-  \frac{\pi}{4}+  \frac{\pi}{4})\beta_{2,l,k}\mathbf{e}^{j\angle\mathbf{H}_{1,l}^H\mathbf{H}_{1,l}}\odot\mathbf{I}_N+
        \frac{\pi\beta_{2,l,k}}{4}
       \mathbf{e}^{j\angle\mathbf{H}_{1,l}^H\mathbf{H}_{1,l}}\odot(\mathbf{1}_N-\mathbf{I}_N)\label{eqsigmavkl},\\&=
        \beta_{2,l,k}\mathbf{I}_N+
        \frac{\pi\beta_{2,l,k}}{4}
       \mathbf{e}^{j\angle\mathbf{H}_{1,l}^H\mathbf{H}_{1,l}}\odot(\mathbf{1}_N-\mathbf{I}_N),%\mathbf{e}^{j\angle\mathbf{b}_{l}\mathbf{b}^H_{l}}\odot\mathbf{e}^{j\angle\mathbf{H}_{1,l}^H\mathbf{H}_{1,l}}\odot
       %\\&= 
        %\beta_{2,l,k}\mathbf{I}_N+
        %\frac{\pi\beta_{2,l,k}}{4}
       %\mathbf{e}^{j\angle\mathbf{b}_{l}\mathbf{b}_{l}^H}\odot(\mathbf{1}_N-\mathbf{I}_N).
        \end{align}
    In step \eqref{eqsigmalemma}, we recall the model of $\mathbf{H}_{1,l}$ in \eqref{eqmodel_H1}. Notice that \eqref{eqsigmavkl} follows from Lemma \ref{lemdad} and using the definitions of mean and variance of a Rayleigh distributed random variable. Furthermore,  $  \boldsymbol{\mu}_{\tilde{\mathbf{v}}_{l}^{{k_l}}}$ is derived as $
        \boldsymbol{\mu}_{\tilde{\mathbf{v}}_{l}^{{k_l}}}=\mathbb{E}[\tilde{\mathbf{v}}_{l}^{{k_l}}]= \mathbb{E}[\text{diag}(|\mathbf{h}_{2,l,k}|)\mathbf{e}^{j\angle\mathbf{H}_{1,l}^H\mathbf{h}_{d,k}}]=\mathbb{E}[\text{diag}(|\mathbf{h}_{2,l,k}|)]\mathbb{E}[\mathbf{e}^{j\angle\mathbf{H}_{1,l}^H\mathbf{h}_{d,k}}]]=\mathbf{0}_{N \times 1}. $
        This follows from the independence between the direct and IRS-user channels, and from the fact that the phase distribution of a circularly symmetric complex normal random variable is uniform leading to $\mathbb{E}[\tilde{\mathbf{v}}_{l}^{{k_l}}]=0$.

        As for the fourth term, recall that $\mathbf{v}_{l}^{k_l}$ \eqref{eqirsphase} in $E_4$ \eqref{eqE4} is the  beamforming vector of non-associated IRS $l$ with respect to user $k$ for which $k_l\neq k$ such that  $\lambda_{l,k}=0$ and $\lambda_{l,k_l}=1$. Its covariance matrix $\boldsymbol{\Sigma}_{\mathbf{v}_{l}^{k_l}} =\mathbf{I}_N$ which can be seen as
      \begin{align}
          \boldsymbol{\Sigma}_{\mathbf{v}_{l}^{k_l}} &=\mathbb{E}[\mathbf{v}_{l}^{k_l}\mathbf{v}_{l}^{k_l^H}]=\mathbb{E}[e^{j \angle{\text{diag}(\mathbf{h}_{2,l,k_l}^H)\mathbf{b}_{l}}}e^{j \angle{\mathbf{a}_{l}^H\mathbf{h}_{d,k_l}}}e^{j \angle{\mathbf{b}_{l}^H\text{diag}(\mathbf{h}_{2,l,k_l})}}e^{j \angle{\mathbf{h}_{d,k_l}^H\mathbf{a}_{l}}}]\\&=\mathbb{E}[e^{j \angle{\text{diag}(\mathbf{h}_{2,l,k_l}^H)\mathbf{b}_{l}}}e^{j \angle{\mathbf{b}_{l}^H\text{diag}(\mathbf{h}_{2,l,k_l})}}]\label{eqsigmavkbarid}
      \end{align}
      If we look at each element in the matrix in       \eqref{eqsigmavkbarid}, we find that the diagonal elements are all ones, and the off-diagonals are all zeros due to independence between  ${h}_{2,l,k_l,n}$ and ${h}_{2,l,k_l,\bar{n}}$ when $n\neq \bar{n}$, so that
\begin{align}
    &\mathbb{E}[[e^{j \angle{\text{diag}(\mathbf{h}_{2,l,k_l}^H)\mathbf{b}_{l}}}e^{j \angle{\mathbf{b}_{l}^H\text{diag}(\mathbf{h}_{2,l,k_l})}}]_{n,\bar{n}}]=\mathbb{E}[e^{j( \angle{{h}_{2,l,k_l,n}^*{b}_{l,n}}-\angle{{h}_{2,l,k_l,\bar{n}}^*{b}_{l,\bar{n}}})}]=0.
\end{align}      

Using the above $\boldsymbol{\Sigma}_{\mathbf{v}_{l}^{k_l}} =\mathbf{I}_N$, we obtain
        \begin{align}
       E_4&=\mathbb{E}[\sum_{l=1}^{L}\sum_{\bar{l}=1}^{L}(1-\lambda_{l,k})(1-\lambda_{\bar{l},k})\mathbf{v}_{l}^{k_l^H}  \mathbf{H}_{0,l,k}^H\mathbf{H}_{0,\bar{l},k}\mathbf{v}_{\bar{l}}^{{k_{\bar{l}}}}],\\ \label{ann_eq} &=
       \sum_{l=1}^{L}(1-\lambda_{l,k})\mathbb{E}[\mathbf{v}_{l}^{k_l^H}  \mathbf{H}_{0,l,k}^H\mathbf{H}_{0,{l},k}\mathbf{v}_{{l}}^{k_l}]= \sum_{l=1}^{L}(1-\lambda_{l,k})\beta_{2,l,k}tr(\mathbf{A}),   \end{align}
        with $\mathbf{A}$ defined as in \eqref{eqA}. Note that the derivation of \eqref{ann_eq} follows using similar steps as done for  $E_3$ in \eqref{eqe3all}. The fifth term $E_5$ is equal to zero due to the zero conditional expectation of  $\mathbf{h}_{d,k}$ given $\mathbf{h}_{d,k_l},\mathbf{h}_{2,l,k_l},\mathbf{h}_{2,l,k}$ where $k_l\neq k$ as follows
  \begin{align}
       E_5&=
     % \mathbb{E}[ 2\sum_{l=1}^{L}(1-\lambda_{l,k})\mathbf{v}_{l}^{k_l}^{^H}  \mathbf{H}_{0,l,k}^H\mathbf{h}_{d,k}\},\\&=
     2\sum_{l=1}^{L}(1-\lambda_{l,k})\mathbb{E}[\mathbf{v}_{l}^{k_l^H}  \mathbf{H}_{0,l,k}^H   \mathbb{E}[\mathbf{h}_{d,k}|\mathbf{h}_{d,k_l},\mathbf{h}_{2,l,k_l},\mathbf{h}_{2,l,k}]]=0
        \end{align}
       Finally, the sixth term $E_6$ is zero using similar arguments as $E_5$. %follows
  %\begin{align}
   %    E_6&=
      %\mathbb{E}[ 2\sum_{l=1}^{L}\sum_{\bar{l}\neq l}^{L}(1-\lambda_{l,k})\lambda_{\bar{l},k}\mathbf{v}_{l}^{k_l}^{H}  \mathbf{H}_{0,l,k}^H\mathbf{H}_{0,\bar{l},k}\mathbf{v}_{\bar{l}}^{{k_{\bar{l}}}}], \\&= 2\sum_{l=1}^{L}\sum_{\bar{l}\neq l}^{L}(1-\lambda_{l,k})\lambda_{\bar{l},k}\mathbb{E}[\mathbf{v}_{l}^{k_l}^{H}   \mathbf{H}_{0,l,k}^H\mathbf{H}_{0,\bar{l},k}\mathbf{v}_{\bar{l}}^{{k_{\bar{l}}}}], \\&=
     % 2\sum_{l=1}^{L}\sum_{\bar{l}\neq l}^{L}(1-\lambda_{l,k})\lambda_{\bar{l},k}\mathbb{E}[\mathbf{v}_{l}^{k_l}^{H}  \mathbf{H}_{0,l,k}^H\mathbf{H}_{1,\bar{l}}|\text{diag}(\mathbf{h}_{2,l,k})| %\times\\ \nonumber
      %\mathbb{E}_{\mathbf{h}_{d,k}}[\mathbf{e}^{j\angle\mathbf{H}_{1,l}^H\mathbf{h}_{d,k}}|\mathbf{h}_{d,t},\mathbf{h}_{2,l,k},\mathbf{h}_{2,l,t}]]=0.% 2\sum_{l=1}^{L}\sum_{\bar{l}\neq l}^{L}(1-\lambda_{l,k})\lambda_{\bar{l},k}\mathbf{v}_{l}^{k_l}^{k_l^H}
       % \end{align} 
       % Thus, we complete the proof and find that collecting these terms yields the result in Lemma \eqref{lemmanormhk2}, stated here for completeness.
  %      \begin{align}
 %\mathbb{E}[\|{\mathbf{h}}_{k}(\boldsymbol{\lambda}_{k})\|^2]&=M\beta_{d,k}+\sum_{l=1}^{L}\lambda_{l,k}\bigg(\sqrt{M}{N}\sqrt{\beta_{1,l}\beta_{2,l,k}\beta_{d,k}}\frac{\pi}{2}\\ \nonumber &+tr(\mathbf{H}_{1,l}^H\mathbf{H}_{1,{l}}\boldsymbol{\Sigma}_{\tilde{\mathbf{v}}_{l}^{k_l}})-\beta_{2,l,k}tr(\mathbf{H}_{1,l}^H\mathbf{H}_{1,{l}}) \bigg)+ \sum_{l=1}^{L}\beta_{2,l,k}tr(\mathbf{H}_{1,l}^H\mathbf{H}_{1,{l}}).
 %   \end{align}

\section{Proof of Lemma \ref{lemmaRk}}

\label{AppendDerivationRK}
In Lemma \ref{lemmaRk}, we defined      $\mathbf{R}_{k}$ as the covariance matrix of $\mathbf{h}_k(\boldsymbol{\lambda}_k)$. In this appendix, we derive  this matrix. Note that
           \begin{align}
           \label{eqrkexp}
        \mathbf{R}_{k}= \mathbf{R}_{k,1}+\mathbf{R}_{k,2}+\mathbf{R}_{k,3}+\mathbf{R}_{k,4}+\mathbf{R}_{k,5}+\mathbf{R}_{k,6},
        \end{align}
        where
           \begin{align}
        &\mathbf{R}_{k,1}=\mathbb{E}[\mathbf{h}_{d,k}\mathbf{h}_{d,k}^H]=\beta_{d,k}\mathbf{I}_M,\hspace{0.2in}\mathbf{R}_{k,2}=\mathbb{E}[2\sum_{l=1}^{L}\lambda_{l,k}\mathbf{h}_{d,k} \mathbf{v}_{l}^{{k_l}^H}   \mathbf{H}_{0,l,k}^H],\\&\mathbf{R}_{k,3}=\mathbb{E}[\sum_{l=1}^{L}\sum_{\bar{l}=1}^{L}\lambda_{l,k}\lambda_{\bar{l},k} \mathbf{H}_{0,l,k}\mathbf{v}_{l}^{k_l} \mathbf{v}_{\bar{l}}^{{k_{\bar{l}}}^H} \mathbf{H}_{0,\bar{l},k}^H],\\&\mathbf{R}_{k,4}=\mathbb{E}[\sum_{l=1}^{L}\sum_{\bar{l}=1}^{L}(1-\lambda_{l,k})(1-\lambda_{\bar{l},k})  \mathbf{H}_{0,l,k}\mathbf{v}_{l}^{k_l}\mathbf{v}_{\bar{l}}^{k_{\bar{l}}^H}\mathbf{H}_{0,\bar{l},k}^H],
        \\&\mathbf{R}_{k,5}=\mathbb{E}[2\sum_{l=1}^{L}(1-\lambda_{l,k}) \mathbf{H}_{0,l,k}\mathbf{v}_{l}^{k_l}\mathbf{h}_{d,k}^H ]=\mathbf{0}_{M\times M},\\&\mathbf{R}_{k,6}=\mathbb{E}[2\sum_{l=1}^{L}\sum_{\bar{l}=1}^{L}(1-\lambda_{\bar{l},k})\lambda_{{l},k}\mathbf{H}_{0,\bar{l},k}\mathbf{v}_{\bar{l}}^{{k_{\bar{l}}}}\mathbf{v}_{l}^{{k_l}^H}  \mathbf{H}_{0,l,k}^H]=\mathbf{0}_{M\times M},.
           \end{align}
      %We find that
       % \begin{align}
        %    \mathbf{R}_{k,1}=\mathbb{E}[\mathbf{h}_{d,k}\mathbf{h}_{d,k}^H]=\beta_{d,k}\mathbf{I}_M
        %\end{align}
        %which comes from the uncorrelated Rayleigh model for $\mathbf{h}_{d,k}$ defined in \eqref{model_hdk}.

%If we let $R_{k,2}=tr({E_2})/M$ , which is the second matrix in the sum for $R_k$ where all the other terms are found and verified earlier. We achieve an upper bound on $R_2$ that is a tight fit due to the large system dimensions. All the terms check out except some terms. The derivation is shown below
In appendix \ref{appendixIEEEproofchannelgainsq}, we show that $\mathbf{R}_{k,5}$ and $\mathbf{R}_{k,6}$ are zero matrices. To compute $\mathbf{R}_{k,2}$, we use the definition of $\mathbf{H}_{0,l,k}^H$ in \eqref{eqholk} to write  
\begin{align}
     \mathbf{R}_{k,2}=&\mathbb{E}[2\sum_{l=1}^{L}\lambda_{l,k}\mathbf{h}_{d,k} \mathbf{v}_{l}^{{k_l}^H} \sqrt{\beta_{1,l}}\text{diag}(\mathbf{h}_{2,l,k}^H)\mathbf{b}_{l}\mathbf{a}_{l}^H].
\end{align}
Its element form can be written as follows
\begin{align}
     [{\mathbf{R}}_{k,2}]_{m,\bar{m}}=&2\sqrt{\beta_{1,l}}\sum_{l=1}^{L}\lambda_{l,k}\mathbb{E}[h_{d,k,m} \mathbf{v}_{l}^{{k_l}^H} \text{diag}(\mathbf{h}_{2,l,k}^H)\mathbf{b}_{l}{a}_{l,\bar{m}}^*],
\end{align}
where ${a}_{l,\bar{m}}^* $ is  the $\bar{m}^{th}$ element in $\mathbf{a}_{l}^H$ and $h_{d,k,m}$ is the ${m}^{th}$ element in $\mathbf{h}_{d,k}$. We can further simplify by using the optimized reflection vector in \eqref{eqirsphase} where $k_l=k$ to obtain
\begin{align}
     [{\mathbf{R}}_{k,2}]_{m,\bar{m}}=&2\sqrt{\beta_{1,l}}\sum_{l=1}^{L}\lambda_{l,k}\mathbb{E}[{h}_{d,k,m} \sum_{n=1}^N e^{j\angle{\mathbf{h}_{d,k}^H\mathbf{a}_{l}}}e^{j\angle{b_{l,n}^*{h}_{2,l,k,n} }}   {h}_{2,l,k,n}^*{b}_{l,n}{a}_{l,\bar{m}}^*].
\end{align}
Here, ${h}_{2,l,k,n}$ is the ${n}^{th}$ element in $\mathbf{h}_{2,l,k}$, and ${b}_{l,n}$ is the ${n}^{th}$ element in $\mathbf{b}_{l}$. %and $\mathbf{h}_{1,l,n}\in\mathbb{C}^{M \times 1}$ is the ${n}^{th}$ column in $\mathbf{H}_{1,l}$.
Thus, the phases cancel to yield
\begin{align}
     [{\mathbf{R}}_{k,2}]_{m,\bar{m}}=&2\sqrt{\beta_{1,l}}\sum_{l=1}^{L}\lambda_{l,k}\sum_{n=1}^N\sqrt{\frac{\pi\beta_{2,l,k}}{4}}{a}_{l,\bar{m}}^*\mathbb{E}\left[{h}_{d,k,m} e^{j\angle{\mathbf{h}_{d,k}^H\mathbf{a}_{l} }}\right]\\&=2\sqrt{\beta_{1,l}}\sum_{l=1}^{L}\lambda_{l,k}\sum_{n=1}^N\sqrt{\frac{\pi\beta_{2,l,k}}{4}}e^{j\angle{{a}_{l,\bar{m}}^*{a}_{l,m}}}\mathbb{E}\left[|{h}_{d,k,m} | e^{j\angle{(1+\frac{\sum_{i\neq m}^M{h}_{d,k,i}^*{a}_{l,i})}{{h}_{d,k,m}^*{a}_{l,m}})}}\right]
\\&=2\sqrt{\beta_{1,l}}N\sum_{l=1}^{L}\lambda_{l,k}\sqrt{\frac{\pi\beta_{2,l,k}}{4}}e^{j\angle{{a}_{l,\bar{m}}^*{a}_{l,m}}}C_m,
\end{align}
where $C_m=\mathbb{E}[|{h}_{d,k,m} | e^{j\angle{(1+\frac{\sum_{i\neq m}^M{h}_{d,k,i}^*{a}_{l,i})}{{h}_{d,k,m}^*{a}_{l,m}})}}]=\sqrt{\frac{\pi\beta_{d,k}}{4M}}$ by noting that $E_2=tr(\mathbf{R}_{k,2})$ where $E_2$ is defined in \eqref{eqE2final}. Therefore we obtain
 \begin{align}
     [\mathbf{R}_{k,2}]_{m,\bar{m}}=&2\sum_{l=1}^{L}\lambda_{l,k}\sqrt{\beta_{1,l}\beta_{2,l,k}\beta_{d,k}}N{\frac{\pi}{4\sqrt{M}}}e^{j\angle{{a}_{l,\bar{m}}^*{a}_{l,m}}}.
\end{align}
%When $M=1$, this expression reduces to 
% \begin{align}
 %    [\mathbf{R}_{k,2,1}]_{1,1}=&N{\frac{\pi}{4}}
%\end{align}
To compute $\mathbf{R}_{k,3}$, we have $k_l=k$ and we express it as
\begin{align}
\label{eqrk3sum}
\mathbf{R}_{k,3}&=\mathbb{E}[ \sum_{l=1}^{L}\sum_{\bar{l}=1}^{L}\lambda_{l,k}\lambda_{\bar{l},k} \mathbf{H}_{0,l,k}\mathbf{v}_{l}^{k_l} \mathbf{v}_{\bar{l}}^{{k_{\bar{l}}}^H} \mathbf{H}_{0,\bar{l},k}^H]=\sum_{l=1}^{L}\lambda_{l,k} \mathbb{E}[\mathbf{H}_{0,l,k}\mathbf{v}_{l}^{k_l} \mathbf{v}_{l}^{{k_l}^H} \mathbf{H}_{0,{l},k}^H]\\ \label{Rk,3}
  &=\sum_{l=1}^{L}\lambda_{l,k}{\beta_{1,l}}\mathbf{H}_{1,l} \mathbb{E}[\text{diag}(|\mathbf{h}_{2,l,k}|) \mathbf{e}^{j\angle\mathbf{H}_{1,l}^H\mathbf{h}_{d,k}}\mathbf{e}^{j\angle\mathbf{h}_{d,k}^H\mathbf{H}_{1,l}}\text{diag}(|\mathbf{h}_{2,l,k}^H|)]\mathbf{H}_{1,l}^H
%=\sum_{l=1}^{L}\lambda_{l,k}{\beta_{1,l}}\mathbf{a}_{l} \mathbf{a}_{l}^H\tilde{\mathbf{R}}_{k,3}
\\&\label{eqlastequiv}=\sum_{l=1}^{L}\lambda_{l,k}{\beta_{1,l}}\mathbf{H}_{1,l}\boldsymbol{\Sigma}_{\tilde{\mathbf{v}}_{l}^{{k_l}}}\mathbf{H}_{1,l}^H
\end{align}

The summation in \eqref{eqrk3sum} reduces due to independence between the different IRS-users channels.    The last equivalence \eqref{eqlastequiv} is obtained following the result for $\boldsymbol{\Sigma}_{\tilde{\mathbf{v}}_l^{k_l}}$ defined in \eqref{eqsigmavkfirst}. Finally, %Using  $\mathbf{v}_{l}^{k_l}=e^{j\angle{\text{diag}(\mathbf{h}^H_{2,l,k})\mathbf{b}_l}}e^{j\angle{\mathbf{a}_l^H\mathbf{h}_{d,k}}}$, we can find an expression for $\tilde{\mathbf{R}}_{k,3}$ as follows
       % \begin{align}
        %  \tilde{\mathbf{R}}_{k,3}&=  \mathbb{E}[\mathbf{b}_{l}^H\text{diag}(\mathbf{h}_{2,l,k})\mathbf{v}_{l}^{k_l} \mathbf{v}_{l}^{{k_l}^H}\text{diag}(\mathbf{h}_{2,l,k}^H) \mathbf{b}_{l}],
         %=\sum_{n=1}^{N}\sum_{\bar{n}=1}^{N}\mathbb{E}[{b}_{l,n}^*{h}_{2,l,k,n}{v}_{l,n}^{{k}} {v}_{l,\bar{n}}^{{k}^*}{h}_{2,l,k,\bar{n}}^*{b}_{l,\bar{n}}],
         % \end{align}
          %where $b_{l,n}^*,{h}_{2,l,k,n},$ and ${v}_{l,n}^{{k}}$ are elements in 
   %  $\mathbf{b}_{l}^H,\mathbf{h}_{2,l,k},$ and $\mathbf{v}_{l}^{k_l}$, respectively.
     %Note that we can decompose this double sum as
        %   \begin{align}
        %   \label{eqrk3sums}
         % \tilde{\mathbf{R}}_{k,3}&=  \sum_{n=1}^{N}\mathbb{E}[{b}_{l,n}^*{h}_{2,l,k,n}{v}_{l,n}^{k} {v}_{l,n}^{{k}^*}{h}_{2,l,k,n}^*{b}_{l,n}]+\sum_{n=1}^{N}\sum_{\bar{n}\neq n}^{N}\mathbb{E}[{b}_{l,n}^*{h}_{2,l,k,n}{v}_{l,n}^{{k}} {v}_{l\bar{n}}^{k^*}{h}_{2,l,k,\bar{n}}^*{b}_{l,\bar{n}}].
         % \end{align}
          %    The first sum yields $N\mathbb{E}[|{h}_{2,l,k}(n)|^2]=N Var(|{h}_{2,l,k}(n)|)+N\mathbb{E}[|{h}_{2,l,k}(n)|]^2=N(1-\frac{\pi}{4})\beta_{2,l,k}+N\frac{\pi}{4}\beta_{2,l,k}$. Putting it all together, $N\mathbb{E}[|{h}_{2,l,k}(n)|^2]=N\beta_{2,l,k}$. 
          
         % The argument of the second sum is  ${b}_{l}^*({n}){b}_{l}({n})\mathbb{E}[|{h}_{2,l,k}(n)|]\mathbb{E}[|{h}_{2,l,k}(\bar{n})|]={b}_{l}^*({n}){b}_{l}({n})\beta_{2,l,k}\frac{\pi}{4}$. 
       
       % \begin{align}
%\mathbf{R}_{k,3}
 %        &=\sum_{l=1}^{L}\lambda_{l,k} \mathbf{H}_{1,l} \boldsymbol{\Sigma}_{\tilde{\mathbf{v}}_l^{k_l}}\mathbf{H}_{1,l}^H
      %   \end{align}
 \begin{align}
\mathbf{R}_{k,4}&=\mathbb{E}[ \sum_{l=1}^{L}\sum_{\bar{l}=1}^{L}(1-\lambda_{l,k})(1-\lambda_{\bar{l},k})  \mathbf{H}_{0,l,k}\mathbf{v}_{l}^{k_l} \mathbf{v}_{\bar{l}}^{k_{\bar{l}}^H} \mathbf{H}_{0,\bar{l},k}^H]\\
  &=\sum_{l=1}^{L}\beta_{1,l}(1-\lambda_{l,k}) \mathbb{E}[\mathbf{a}_{l}\mathbf{b}_{l}^H\text{diag}(\mathbf{h}_{2,l,k})\mathbf{v}_{l}^{k_l} \mathbf{v}_{l}^{k_l^H}\text{diag}(\mathbf{h}_{2,l,k}^H) \mathbf{b}_{l}\mathbf{a}_{l}^H]
  =\sum_{l=1}^{L}\beta_{1,l}(1-\lambda_{l,k}) \tilde{\mathbf{R}}_{k,4}
\end{align}
The summation reduces due to independence between the different IRS-users channels.
        Using
        $\mathbf{v}_{l}^{k_l}$ in \eqref{eqirsphase}, for $k_l \neq k$ and defining $\tilde{\mathbf{R}}_{k,4}$ we get
        \begin{align}
          \tilde{\mathbf{R}}_{k,4}&=  \mathbb{E}[\mathbf{a}_{l}\mathbf{b}_{l}^H\text{diag}(\mathbf{h}_{2,l,k})e^{j\angle{\text{diag}(\mathbf{h}^H_{2,l,k_l})\mathbf{b}_l}} e^{j\angle{\mathbf{b}_l}^H\text{diag}(\mathbf{h}_{2,l,k_l})}\text{diag}(\mathbf{h}_{2,l,k}^H) \mathbf{b}_{l}\mathbf{a}_{l}^H]\\&=\mathbf{a}_{l}\mathbf{b}_{l}^H   \mathbb{E}_{\mathbf{h}_{2,l,k}}[  \mathbb{E}[\text{diag}(\mathbf{h}_{2,l,k})e^{j\angle{\text{diag}(\mathbf{h}^H_{2,l,k_l})\mathbf{b}_l}} e^{j\angle{\mathbf{b}_l}^H\text{diag}(\mathbf{h}_{2,l,k_l})}\text{diag}(\mathbf{h}_{2,l,k}^H)|\mathbf{h}_{2,l,k_l}]]\mathbf{b}_{l}\mathbf{a}_{l}^H, \\&=\beta_{2,l,k} \|\mathbf{b}_{l}\|^2\mathbf{a}_{l}\mathbf{a}_{l}^H.
\end{align}
  Notice that $\mathbb{E}[\text{diag}(\mathbf{h}_{2,l,k})e^{j\angle{\text{diag}(\mathbf{h}^H_{2,l,k_l})\mathbf{b}_l}} e^{j\angle{\mathbf{b}_l}^H\text{diag}(\mathbf{h}_{2,l,k_l})}\text{diag}(\mathbf{h}_{2,l,k}^H)|\mathbf{h}_{2,l,k}]$ can be seen as random diagonal matrix $\text{diag}(\mathbf{h}^H_{2,l,k})$ multiplied by a deterministic matrix $e^{j\angle{\text{diag}(\mathbf{h}^H_{2,l,k_l})\mathbf{b}_l}} e^{j\angle{\mathbf{b}_l}^H\text{diag}(\mathbf{h}_{2,l,k_l})}$ multiplied by the same diagonal matrix. Therefore, we have the right set-up to apply Lemma \ref{lemdad} and get the following: the off-diagonals are equal to zero, while the diagonal terms are equal to $\beta_{2,l,k}$. Thus, we achieve a scaled identity for this expectation term which simplifies our expression when plugging $\tilde{\mathbf{R}}_{k,4}$ for $\mathbf{R}_{k,4}$ to obtain
 \begin{align}
\mathbf{R}_{k,4}&=\sum_{l=1}^{L}(1-\lambda_{l,k}) \beta_{1,l}\beta_{2,l,k}\|\mathbf{b}_{l}\|^2\mathbf{a}_{l}\mathbf{a}_{l}^H=
\sum_{l=1}^{L}(1-\lambda_{l,k}) {\beta_{2,l,k}}\mathbf{H}_{1,l}\mathbf{H}_{1,{l}}^H \end{align}
           
         Thus, we have completed the proof for Lemma \ref{lemmaRk}.%, and the expression for $        \mathbf{R}_{k}$ is given as
           %   \begin{align}
        %\mathbf{R}_{k}=&\beta_{d,k}\mathbf{I}_M+\sum_{l=1}^{L}\lambda_{l,k}\bigg(2\sqrt{\beta_{1,l}\beta_{2,l,k}\beta_{d,k}}\frac{N\pi}{4\sqrt{M}} \mathbf{e}^{j\angle\mathbf{H}_{1,l}\mathbf{H}_{1,l}^H}\\ \nonumber &+ \mathbf{H}_{1,l}
       %\boldsymbol{\Sigma}_{\tilde{\mathbf{v}}_{l}^{k_l}} \mathbf{H}_{1,l}^H- \beta_{2,l,k}\mathbf{H}_{1,l}\mathbf{H}_{1,{l}}^H\bigg)+\sum_{l=1}^{L}\beta_{2,l,k}\mathbf{H}_{1,l}\mathbf{H}_{1,{l}}^H.
    %       \end{align}  

       %\appendix

   \end{appendices}
     \bibliographystyle{IEEEtran}
\bibliography{bibli}
   
\end{document}